\documentclass[aip,pop,reprint]{revtex4-1}
\usepackage{mathptmx}
\usepackage[T1]{fontenc}
\usepackage[utf8]{inputenc}
\usepackage{color}
\usepackage{xcolor}
\usepackage{units}
\usepackage{mathrsfs}
\usepackage{mathtools}
\usepackage{bm}
\usepackage{amsmath}
\usepackage{amssymb}
\usepackage{graphicx}
\usepackage{cancel}
\usepackage[pdfusetitle,
 bookmarks=true,bookmarksnumbered=false,bookmarksopen=false,
 breaklinks=false,pdfborder={0 0 1},backref=false,colorlinks=true]
 {hyperref}
\hypersetup{
 pdfborderstyle=,linkcolor=blue,citecolor=blue}

\makeatletter
\usepackage{etoolbox}
\allowdisplaybreaks

\def\@email#1#2{%
 \endgroup
 \patchcmd{\titleblock@produce}
  {\frontmatter@RRAPformat}
  {\frontmatter@RRAPformat{\produce@RRAP{*#1\href{mailto:#2}{#2}}}\frontmatter@RRAPformat}
  {}{}
}%

\makeatother

\begin{document}
\title{Wave instabilities in anisotropic regularized Kappa plasmas: New plasma dispersion functions and numerical validation}
\author{Rudi Gaelzer}
\affiliation{Instituto de Física, Universidade Federal do Rio Grande do Sul, CP
15051, 91501-970, Porto Alegre, RS, Brazil}
\email{rudi.gaelzer@ufrgs.br}

\author{Dustin L. Schröder}
\affiliation{Institut für Theoretische Physik IV, Ruhr-Universität Bochum, Universitätsstrasse
150, 44780 Bochum, Germany}
\author{Marian Lazar}
\affiliation{Institut für Theoretische Physik IV, Ruhr-Universität Bochum, Universitätsstrasse
150, 44780 Bochum, Germany}
\affiliation{Centre for Mathematical Plasma Astrophysics, Department of Mathematics,
KU Leuven, Belgium}
\author{Horst Fichtner}
\affiliation{Institut für Theoretische Physik IV, Ruhr-Universität Bochum, Universitätsstrasse
150, 44780 Bochum, Germany}
\begin{abstract}
Plasmas in space and astrophysical environments are frequently found in (quasi-)stationary states not in thermal equilibrium, indicated by velocity distribution functions (VDFs) featuring nonthermal characteristics, such as Lorentzian high-energy tails and kinetic (temperature and strahl/beam) anisotropies. For decades, in-situ observations of electrons and ions in the solar corona, solar wind or planetary magnetospheres have been modelled using so-called Kappa VDFs, whose standard form, however - depending on the considered VDF moment - imposes stringent constraints on the minimum value of the Kappa power-law parameter. Such limitations have posed difficulties on the derivation of a fully-consistent hydrodynamic formulation of suprathermal plasmas. To improve this situation, the recently proposed regularized Kappa VDF avoids strict constraints on the Kappa parameter by introducing an exponential, regularizing cut-off. In this work we assume that the plasma particles are described by regularized-Kappa VDFs that feature anisotropies not only in their thermal velocity spread parameters but also in their exponential cut-offs. The dielectric tensor components for waves propagating along a magnetic field are derived, in terms of new plasma dispersion functions. The resulting dispersion equations are solved with two complementary techniques, namely by using the analytical expressions derived here and by employing the ALPS code. This is done for physical parameters that give rise to electromagnetic and electrostatic instabilities. Both approaches show excellent agreement and the results demonstrate how an anisotropy in the regularizing cut-off of VDFs can serve, depending on an instability's nature, either as a stabilization factor or as an additional free energy source.
\end{abstract}

\maketitle

\section{Introduction}

Space plasmas are typically hot and dilute, and therefore nearly collisionless. 
As a result, these plasmas are not in thermodynamic equilibrium, which in-situ observations attribute primarily to kinetic anisotropies and suprathermal populations \citep{Stverak-etal-2008, Wilson-etal-2019a, Wilson-etal-2019b}. 
Interest for understanding such observational evidences and their impact in the solar wind and planetary environments is currently extending to the solar corona, driven by ongoing exploration through the Parker Solar Probe \citep{Halekas-etal-2021, Abraham-etal-2022} and Solar Orbiter missions \citep{Bercic-etal-2021, Coburn-etal-2024}.
It is expected that the dynamics and properties of such weakly collisional plasmas are governed by wave fluctuations, including the kinetic instabilities induced by the anisotropies of the velocity distributions \citep{Verscharen-etal-2022}.
These instabilities can be stimulated by the suprathermal populations \citep{Lazar-etal-2022, Shaaban-etal-2024}, and the resulting plasma fluctuations interact back with the particles, limiting their anisotropies \citep{Stverak-etal-2008, Verscharen-etal-2019}.

Despite such nonlinear coupling of the plasma fluctuations with the plasma populations, linear dispersion theory remains to be a valuable standard tool\cite{Gary05, Yoon-2017} for the studies of plasma instabilities because predictions from linear theory are essential for interpreting observations \citep{Stverak-etal-2008, Shaaban-etal-2021}, for calibrating numerical simulations \citep{Lopez-etal-2021}, and for developing extended quasi-linear theories \citep{Yoon-2017, Shaaban-etal-2021}.
This involves determining the nature of the waves, frequency ranges, and unstable wave numbers, as well as identifying the fastest-growing modes expected to drive the resulting enhanced waves and the relaxation of anisotropies.
Realistic predictions require velocity distribution functions (VDFs) that closely reproduce in-situ observations.
Thus, the observed VDFs of plasma populations (electrons and protons/ions) reveal not only temperature anisotropies and beam (or drifting) populations, but also suprathermal tails at high energies. 
This led to the introduction of so-called Kappa VDFs \citep{Olbert-1968, Vasyliunas-1968, Scherer-etal-2017} and their anisotropic forms \citep{Summers-Thorne-1991, Scherer+20/07, Scherer-etal-AA2022} as upgrades to idealized Maxwellian models, capable of capturing both the quasi-thermal core at low energies and the suprathermal tails; for recent advancements in these models see Ref.~\onlinecite{Lazar-Fichtner-2021}.

Despite these efforts, the results may be affected by flaws in empirical Kappa models, such as the restriction in defining the moments and in the hydrodynamic macro-modeling \citep{Husidic-etal-2021} to sufficiently high values of the parameter $\kappa$, or the undesired non-physical effects of superluminal particles with speeds higher than the speed of light unavoidably occuring in a non-relativistic model \citep{Scherer-etal-ApJ2019}. To overcome these and other shortcomings,  isotropic \citep{Scherer-etal-2017} and anisotropic \citep{Scherer+20/07} regularized Kappa distributions (RKDs) have been proposed, which in addition to a power-law part exhibit an exponential (Maxwellian-like) cut-off to limit an unrealistic extent of suprathermal tails. Adding such cut-off significantly complicates the analysis, making it harder to analytically derive the dielectric properties essential for studying waves and instabilities. Nonetheless, RKDs are gaining ground in many applications \citep{Husidic-etal-2022, Scherer-etal-AA2022, Hau-etal-2023, El-Tantawy-etal-2024, Gaelzer-Ziebell-2025}, also motivating advances in the analysis of wave dispersion properties with numerical and even theoretical tools \citep{Husidic-etal-2020, Gaelzer+24/07, Hau-etal-2024, Schroeder+25/03, Schroeder+25/07}.

However, the anisotropic RKDs were, so far, mostly employed under the assumption of an isotropic cut-off, i.e.\ a common regularization parameter of the distribution function along and perpendicular to the magnetic field. Given that this does not exploit the full flexibility to model an anisotropy of an RKD, this is referred to as the  \textit{semi-anisotropic} case. 

In the present work, we address the more general case of a \textit{fully anisotropic} RKD, which, in addition to anisotropic 'thermal' speeds, also takes into account different cut-off parameters in the principal directions, parallel ($\parallel$) and perpendicular ($\perp$) to the magnetic field, so that the temperature anisotropy is also determined by such anisotropic cut-off. Such fully anisotropic RKD has so far only been considered in a fully numerical analysis of electromagnetic instabilities\citep{Husidic-etal-2020}\@. In fact, most of the existing dispersion and stability analyses of RKD plasmas rely on numerical solvers \citep{Husidic-etal-2020, Schroeder+25/03, Schroeder+25/07} due to the lack of adequate analytical expressions, such as the plasma dispersion functions, i.e., integral functions and, implicitly, the dielectric tensor components specific to RKD plasmas. However, ensuring the reliability of numerical solvers demands extensive validation across both standard and non-standard distributions and their analytical result \citep{ GaelzerZiebell16/02, Gaelzer+16/06, Verscharen_2018, Lopez_Shaaban_Lazar_2021}. 

Motivated by these needs, in the present paper we extend the semi-analytic approach by deriving the susceptibility tensor components for parallel-propagating waves in a plasma modelled by fully anisotropic RKD distribution functions.  These components are written in terms of new plasma dispersion functions, whose definitions and mathematical properties are also discussed.
The mathematical formalism thus developed is subsequently implemented in a specific computer code that numerically solves the dispersion equation.  In order to verify the results obtained by the semi-analytic approach, they will be compared to results obtained from an independently developed numerical dispersion solver, the ALPS code\cite{Verscharen_2018}\@.
As we shall show, the results from both the semi-analytic and the fully numerical approaches fully agree for all cases considered and so this work also serves as an example of mutual corroboration test between two different approaches to the analysis of wave propagation and linear wave-particle interaction in a magnetized plasma.
This way, we provide a most flexible and general tool for the analysis of kinetic instabilities and, at the same time, 
useful test cases for the validation of fully numerical approaches. 

The paper is organized as follows. In section~\ref{sec:Theoretical-formulation} the dispersion theory based on a fully anisotropic RKD is worked out analytically by deriving anisotropic dispersion functions, and these are implemented in the components of susceptibility tensor for parallel propagation.
Then, sections~\ref{sec:Semi-anisotropic} and~\ref{sec:Fully-anisotropic} contain the applications describing electromagnetic instabilities driven by anisotropic electrons with semi- and fully anisotropic RKDs. 
Section~\ref{sec:Semi-anisotropic} contains the comparisons with results from a recently published work, which were obtained employing solely the ALPS code and for the particular case of a single cut-off parameter.  A detailed analysis of this case serves as an important and necessary initial test of our doubly-pronged approach.
Subsequently, section ~\ref{sec:Fully-anisotropic} extends the mutual corroboration tests to the more general case of anisotropic cut-off parameters.
Finally, a summary of the results and the primary conclusions are presented in  section~\ref{sec:Summary}.

\section{Theoretical formulation}\label{sec:Theoretical-formulation}

In this section, the basic formalism employed in our work will be
introduced.

\subsection{The generalized anisotropic RKD with drift}

The model VDF to be employed in this work extends both the generalized
form introduced by Refs. \onlinecite{Scherer+20/07,Gaelzer+24/07},
and the drift-anisotropic distribution employed by Ref. \onlinecite{Schroeder+25/07},
namely 
\begin{align}
f_{\kappa,\boldsymbol{\theta}}^{(\eta,\zeta,\boldsymbol{\mu})}\left(v_{\parallel},v_{\perp}\right) & =N_{\kappa,\boldsymbol{\theta}}^{(\eta,\zeta,\boldsymbol{\mu})}\left[1+\frac{\left(v_{\parallel}-U_{\parallel}\right)^{2}}{\eta\theta_{\parallel}^{2}}+\frac{v_{\perp}^{2}}{\eta\theta_{\perp}^{2}}\right]^{-\zeta}\nonumber \\
 & \times\exp\left[-\frac{\mu_{\parallel}}{\theta_{\parallel}^{2}}\left(v_{\parallel}-U_{\parallel}\right)^{2}-\frac{\mu_{\perp}}{\theta_{\perp}^{2}}v_{\perp}^{2}\right],\label{eq:RKAP3:RKD-anisotropic-beam-generalized}
\end{align}
where $\boldsymbol{\theta}=\left(\theta_{\parallel},\theta_{\perp}\right)$, $\boldsymbol{\mu}=\left(\mu_{\parallel},\mu_{\perp}\right)$ and $N_{\kappa,\boldsymbol{\theta}}^{(\eta,\zeta,\boldsymbol{\mu})}$
is the normalization constant. 
In \eqref{eq:RKAP3:RKD-anisotropic-beam-generalized}, parameters $\eta$ and $\zeta$ are coupled via $\kappa$ parameter, as in the standard Kappa VDFs recovered for $\mu_\parallel=\mu_\perp =0$, with $\eta = \kappa$ and $\zeta = \kappa +1$. 
Instead, $\mu_{\parallel,\perp}$ are typical dimensionless constants \citep{Husidic-etal-2020, Scherer-etal-AA2022} that do not necessarily depend on \(\kappa \).
Their values balance the reduction of unphysical effects of superluminal particles with the need to retain suprathermal populations in the high-energy tails \citep{Scherer-etal-ApJ2019}.
As it will be shown below, the normalization parameters $\theta_{\parallel,\perp}$ directly link to the temperature components $T_{\kappa, \parallel,\perp}$ (in energy units), as for (drifting) standard bi-Kappa  VDFs~\citep{Gaelzer+24/07,GaelzerZiebell16/02,Gaelzer+16/06}. 
%
\begin{equation}
T_{\kappa, \parallel,\perp} = \frac{\kappa}{\kappa - \nicefrac{3}{2}} \, \frac{m \theta^2_{\parallel,\perp}}{2},
\label{eq:Temperatures_bSKd}
\end{equation}
with $m$ the particle mass.
In this case, the limit $\kappa\to\infty$
will restore the drifting bi-Maxwellian (dbM) VDF 
\begin{equation}
f_{\rm dbM}\left(v_{\parallel},v_{\perp}\right)=\frac{1}{\pi^{3/2}\theta_{\parallel}\theta_{\perp}^{2}}\exp\left(-\frac{\left(v_{\parallel}-U_{\parallel}\right)^{2}}{\theta_{\parallel}^{2}}-\frac{v_{\perp}^{2}}{\theta_{\perp}^{2}}\right),\label{eq:Maxwellian-VDF}
\end{equation}
with temperature components $T_{M, \parallel,\perp} = m \theta^2_{\parallel,\perp}/ 2 < T_{\kappa, \parallel,\perp}$~\citep{Lazar-etal-2015, Lazar-Fichtner-2021}. 

In the isotropic case $\theta_{\parallel}=\theta_{\perp}=\theta$ and $\mu_{\parallel}=\mu_{\perp}=\mu$, distribution \eqref{eq:RKAP3:RKD-anisotropic-beam-generalized} reduces to the model employed by Ref. \onlinecite{Gaelzer+24/07}, whereas in the case $\mu_{\parallel}=\mu_{\perp}=\alpha^{2}$ (with $\theta_{\parallel}\neq\theta_{\perp}$), we obtain the anisotropic distribution adopted by Ref. \onlinecite{Schroeder+25/07}\@. 
A further limiting case without the drift $\left(U_{\parallel}=0\right)$ reproduces the anisotropic VDF adopted by Ref. \onlinecite{Schroeder+25/03}.
%
%
Our form \eqref{eq:RKAP3:RKD-anisotropic-beam-generalized}, therefore, introduces further flexibility by allowing separate choices for the regularization parameters $\left(\mu_{\parallel},\mu_{\perp}\right)$ in either direction.

We will now obtain an equivalent form for the generalized RKD,
suitable for the ensuing evaluation of the dielectric tensor. This
is accomplished via the identity 
\[
\int_{0}^{\infty}\beta^{-\delta}e^{-\gamma\beta}d\beta=\Gamma\left(1-\delta\right)\gamma^{\delta-1}\quad\left(\gamma>0\right),
\]
which renders \eqref{eq:RKAP3:RKD-anisotropic-beam-generalized} as
\begin{multline}
f_{\kappa,\boldsymbol{\theta}}^{(\eta,\zeta,\boldsymbol{\mu})}\left(v_{\parallel},v_{\perp}\right)=\frac{N_{\kappa,\boldsymbol{\theta}}^{(\eta,\zeta,\boldsymbol{\mu})}}{\Gamma\left(\zeta\right)}\\
\times\int_{0}^{\infty}d\beta\,\beta^{\zeta-1}e^{-\beta}e^{-\left(v_{\parallel}-U_{\parallel}\right)^{2}/\phi_{\parallel}^{2}}e^{-v_{\perp}^{2}/\phi_{\perp}^{2}},\label{eq:RKAP3:RKD-anisotropic_beam-superstatistics}
\end{multline}
where 
\[
\phi_{\parallel,\perp}\left(\beta\right)=\frac{\theta_{\parallel,\perp}}{\sqrt{\mu_{\parallel,\perp}+\beta/\eta}}.
\]

We will first use \eqref{eq:RKAP3:RKD-anisotropic_beam-superstatistics}
to obtain the normalization constant $N_{\kappa,\boldsymbol{\theta}}^{(\eta,\zeta,\boldsymbol{\mu})}$,
which is determined from the condition 
\[
\int d^{3}v\,f_{\kappa,\boldsymbol{\theta}}^{(\eta,\zeta,\boldsymbol{\mu})}\left(v_{\parallel},v_{\perp}\right)=1.
\]
Inserting \eqref{eq:RKAP3:RKD-anisotropic_beam-superstatistics},
we can easily evaluate the velocity integrals using a cylindrical
coordinate system, resulting in 
\[
\pi^{3/2}\theta_{\parallel}\theta_{\perp}^{2}\frac{N_{\kappa,\boldsymbol{\theta}}^{(\eta,\zeta,\boldsymbol{\mu})}}{\Gamma\left(\zeta\right)}\int_{0}^{\infty}d\beta\,\frac{\beta^{\zeta-1}e^{-\beta}}{\sqrt{\mu_{\parallel}+\beta/\eta}\left(\mu_{\perp}+\beta/\eta\right)}=1.
\]

The remaining integral is identified with \eqref{eq:RKAP2:IBGD-1}
and with the identity \eqref{eq:RKAP3:I_beta-gen-CalU-gen}, providing
the final expression 
\[
N_{\kappa,\boldsymbol{\theta}}^{(\eta,\zeta,\boldsymbol{\mu})}=\left[\left(\pi\eta\right)^{3/2}\theta_{\parallel}\theta_{\perp}^{2}\mathcal{U}^{\left(\eta,\zeta,\boldsymbol{\mu}\right)}\right]^{-1},
\]
where $\mathcal{U}^{\left(\eta,\zeta,\boldsymbol{\mu}\right)}\equiv\mathcal{U}_{0,0}^{\left(\eta,\zeta,\boldsymbol{\mu}\right)}$,
the latter being given by expressions \eqref{eq:RKAP3:CalU-general}
and \eqref{eq:RKAP3:CalU-gen-series}.
Notice that in the $\mu$-isotropic limit (\emph{i.e.}, $\mu_{\parallel}=\mu_{\perp}=\mu$),
one obtains the known expressions\citep{Scherer+20/07,Gaelzer+24/07}
\begin{align*}
\mathcal{U}^{\left(\eta,\zeta,\mu\right)} &= U\left(\frac{3}{2},\frac{5}{2}-\zeta,\eta\mu\right) \\ 
&= \left(\eta\mu\right)^{\zeta-3/2}U\left(\zeta,\zeta-\frac{1}{2},\eta\mu\right),
\end{align*}
where $U(a,b,z)$ is the second solution of Kummer's equation, known as Tricomi function (see eq. \ref{eq:RKAP2:Tricomi-Integ_rep-1}).

Other important physical properties that can be derived from (\ref{eq:RKAP3:RKD-anisotropic_beam-superstatistics}) are the perpendicular (parallel) kinetic temperatures, evaluated from the second moments of the VDF according to $T_\perp = \frac{1}{2}m \bigl\langle v_\perp^2 \bigr\rangle$ and $T_\parallel = m \bigl\langle (v_\parallel - U_\parallel)^2 \bigr\rangle$\@.  Employing expressions (\ref{eq:RKAP2:IBGD-1}) and (\ref{eq:RKAP3:I_beta-gen-CalU-gen}), one obtains
\begin{align}
T^{(\eta,\zeta,\boldsymbol{\mu})}_{\perp,\kappa,\boldsymbol{\theta}} & =\frac{1}{2}m\theta^{2}_{\perp}\eta\frac{\mathcal{U}^{\left(\eta,\zeta,\boldsymbol{\mu}\right)}_{0,1}}{\mathcal{U}^{\left(\eta,\zeta,\boldsymbol{\mu}\right)}_{0,0}}, & T^{(\eta,\zeta,\boldsymbol{\mu})}_{\parallel,\kappa,\boldsymbol{\theta}} & =\frac{1}{2}m\theta^{2}_{\parallel}\eta\frac{\mathcal{U}^{\left(\eta,\zeta,\boldsymbol{\mu}\right)}_{1,0}}{\mathcal{U}^{\left(\eta,\zeta,\boldsymbol{\mu}\right)}_{0,0}}.
\label{eq:RKAP3:bRKd-Kinetic_temperatures}
\end{align}
These results show that the temperatures of a fully-anisotropic RKD depend not only on the thermal velocity parameters $\theta_{\perp(\parallel)}$ but also on the regularization parameters $\mu_{\perp(\parallel)}$, via the $\mathcal{U}^{\left(\eta,\zeta,\boldsymbol{\mu}\right)}_{\alpha,\beta}$ functions discussed in Appendix \ref{sec:Definitions-and-properties}\@.  
In fact, if $\mu_\perp \neq \mu_\parallel$, a higher degree of anisotropy is possible (compared to the Maxwellian and standard Kappa VDFs), in such a way that $T_\perp / \theta_\perp^2 \neq T_\parallel/\theta_\parallel^2$.
The kinetic temperatures (\ref{eq:Temperatures_bSKd}) for a standard bi-Kappa VDF are the limiting case of the results above when $\mu_{\perp(\parallel)} \to 0$, as one can verify after introducing the limit (\ref{eq:RKAP3:CalU-SKD_limit}) and taking the usual $\eta = \kappa$ and $\zeta = \kappa + 1$.

\subsection{The dielectric tensor and the dispersion equations}

We consider waves propagating parallel to the (homogeneous) magnetic field $\boldsymbol{B}_{0}=B_{0}\hat{\boldsymbol{z}}$ $\left(B_{0}>0\right)$\@.
In this case the only nonzero Cartesian components of the dielectric tensor are $\varepsilon_{xx}=\varepsilon_{yy}=\varepsilon_{1}$, $\varepsilon_{xy}=-\varepsilon_{yx}=\varepsilon_{2}$, and $\varepsilon_{zz}=\varepsilon_{3}$, written as\citep{Brambilla98}
\[
\varepsilon_{i}=\delta_{i1}+\delta_{i3}+\sum_{a}\chi_{i}^{(a)}\quad\left(i=1,2,3\right),
\]
where $\chi_{i}^{(a)}$ is the $i$-th component of the susceptibility tensor associated with particle species/population $a=\left(e,i,\dots\right)$\@.
\begin{subequations}
\label{eq:RKAP2:DT-magnetized-parallel}
For a gyrotropic distribution function $f_{a0}=f_{a0} (v_{\parallel}, v_{\perp})$, the susceptibility components for parallel propagation are\citep{Brambilla98}
\begin{align}
\chi_{1}^{(a)} & =\frac{1}{4}\frac{\omega_{pa}^{2}}{\omega^{2}}\sum_{s=\pm1}\int d^{3}v\frac{v_{\perp}\mathcal{L}f_{a0}}{\omega-s\Omega_{a}-k_{\parallel}v_{\parallel}}\label{eq:RKAP2:DTMP-E1}\\
\chi_{2}^{(a)} & =i\frac{1}{4}\frac{\omega_{pa}^{2}}{\omega^{2}}\sum_{s=\pm1}s\int d^{3}v\frac{v_{\perp}\mathcal{L}f_{a0}}{\omega-s\Omega_{a}-k_{\parallel}v_{\parallel}}\label{eq:RKAP2:DTMP-E2}\\
\chi_{3}^{(a)} & =\frac{\omega_{pa}^{2}}{k_{\parallel}}\int d^{3}v\frac{\partial f_{a0}/\partial v_{\parallel}}{\omega-k_{\parallel}v_{\parallel}}.\label{eq:RKAP2:DTMP-E3}
\end{align}
In \eqref{eq:RKAP2:DTMP-E2} -- \eqref{eq:RKAP2:DTMP-E3}, 
we have defined the differential operator\[
\mathcal{L}f_{a0}=\left(\omega-k_{\parallel}v_{\parallel}\right)\frac{\partial f_{a0}}{\partial v_{\perp}}+k_{\parallel}v_{\perp}\frac{\partial f_{a0}}{\partial v_{\parallel}},
\]
where $v_{\parallel}$ and $v_{\perp}$ are, respectively, the parallel and perpendicular (to $\bm{B}_{0}$) components of the particle velocity.
Likewise, $k_{\parallel}$ is the parallel component of the wavevector $\left(k_{\perp}=0\right)$ and $\omega$ is the wave angular frequency. Each particle species/population is characterized by its number density $n_{a}$ and the particle's mass $m_{a}$ and charge $q_{a}$\@. Accordingly, $\omega_{pa}=\sqrt{4\pi n_{a}q_{a}^{2}/m_{a}}$
and $\Omega_{a}=q_{a}B_{0}/m_{a}c$ are the $a$-species plasma and
cyclotron frequencies, with $c$ being the vacuum light speed.%

\end{subequations}

The linearized Vlasov-Maxwell equations leads to the wave
equation for a magnetized plasma that, in the case of parallel propagation
and in the Fourier $\left(\bm{k},\omega\right)$ space, factors into
two branches:\citep{Brambilla98} the dispersion equation for longitudinal
waves 
\begin{gather}
\varepsilon_{3}=1+\sum_{a}\chi_{3}^{(a)}=0,\label{eq:RKAP2:DE-electrostatic}
\end{gather}
and the dispersion equations for electromagnetic waves
\begin{gather}
N_{s}^{2}=1+\sum_{a}\left(\chi_{1}^{(a)}-si\chi_{2}^{(a)}\right)\,\left(s=\pm1\right),\label{eq:RKAP2:DE-RL}
\end{gather}
where $N_{s}=k_{\parallel}c/\omega_{s}$ is the index of refraction of the $s$-th mode. 
For $s=+1$ the equation describes left-handed circularly polarized ($L$) waves $\left(N_{+}=N_{L}\right)$, whereas the equation for $s=-1$ corresponds to right-handed ($R$) waves $\left(N_{-}=N_{R}\right)$.
The dispersion equations \eqref{eq:RKAP2:DE-electrostatic} and \eqref{eq:RKAP2:DE-RL} depend on the evaluation of the components (\ref{eq:RKAP2:DT-magnetized-parallel}a-c) of the susceptibility tensor for all plasma species/populations, and ultimately on the evaluation of the integrals contained therein.

\begin{subequations}
\label{eq:DT-bRKDd}

The susceptibility tensor of a species described by the anisotropic
drifting bi-RKD (hereafter identified by the shortened acronym ``dbRK'') was obtained. 
The evaluation of the integrals is simplified if one employs the form \eqref{eq:RKAP3:RKD-anisotropic_beam-superstatistics} of the VDF, and the susceptibility tensor components can be written as
\begin{align}
\chi_{1}^{(\mathrm{dbRK})} & =\frac{1}{2}\frac{\omega_{pa}^{2}}{\omega^{2}}\sum_{s=\pm1}\left[\varsigma_{0}Z_{\kappa}^{\left(\eta,\zeta,\boldsymbol{\mu}\right)}\left(\varsigma_{s}\right)+\frac{1}{2}\mathcal{Z}_{\kappa,\boldsymbol{\theta}}^{\left(\eta,\zeta,\boldsymbol{\mu}\right)\prime}\left(\varsigma_{s}\right)\right]\\
\chi_{2}^{(\mathrm{dbRK})} & =\frac{i}{2}\frac{\omega_{pa}^{2}}{\omega^{2}}\sum_{s=\pm1}s\left[\varsigma_{0}Z_{\kappa}^{\left(\eta,\zeta,\boldsymbol{\mu}\right)}\left(\varsigma_{s}\right)+\frac{1}{2}\mathcal{Z}_{\kappa,\boldsymbol{\theta}}^{\left(\eta,\zeta,\boldsymbol{\mu}\right)\prime}\left(\varsigma_{s}\right)\right]\\
\chi_{3}^{(\mathrm{dbRK})} & =-\frac{\omega_{pa}^{2}}{k_{\parallel}^{2}\theta_{\parallel}^{2}}Z_{\kappa}^{\left(\eta,\zeta,\boldsymbol{\mu}\right)\prime}\left(\varsigma_{0}\right).
\end{align}
The new dispersion functions $Z_{\kappa}^{\left(\eta,\zeta,\boldsymbol{\mu}\right)}\left(z\right)$
and $\mathcal{Z}_{\kappa,\boldsymbol{\theta}}^{\left(\eta,\zeta,\boldsymbol{\mu}\right)}\left(z\right)$,
as well as their derivatives $Z_{\kappa}^{\left(\eta,\zeta,\boldsymbol{\mu}\right)\prime}=dZ_{\kappa}^{\left(\eta,\zeta,\boldsymbol{\mu}\right)}/dz$
and $\mathcal{Z}_{\kappa,\boldsymbol{\theta}}^{\left(\eta,\zeta,\boldsymbol{\mu}\right)\prime}=d\mathcal{Z}_{\kappa,\boldsymbol{\theta}}^{\left(\eta,\zeta,\boldsymbol{\mu}\right)}/dz$,
are defined and analysed in section \ref{subsec:The-dispersion-functions}\@. Their arguments are ($n=0,\pm1$)
\begin{align*}
\varsigma_{n} & =\xi_{n}-\frac{U_{\parallel}}{\theta_{\parallel}}, & \xi_{n} & =\frac{\omega-n\Omega}{k_{\parallel}\theta_{\parallel}}.
\end{align*}
\end{subequations}

\subsection{The dispersion functions for anisotropic RKD plasmas} \label{subsec:The-dispersion-functions}

As oftentimes happens whenever one endeavours to devise an analytical description for the dispersive properties of waves propagating in (and interacting with) plasmas statistically described by new forms of VDFs, this objective can only be achieved by defining new dispersion functions. 
In this section two new functions (and their derivatives) are defined for dbRK VDFs, and some of their properties are derived.

Thus, the new dispersion function associated to longitudinal waves is %
\begin{multline}
Z_{\kappa}^{\left(\eta,\zeta,\boldsymbol{\mu}\right)}\left(z\right)=\frac{\eta^{-3/2}}{\mathcal{U}^{\left(\eta,\zeta,\boldsymbol{\mu}\right)}}\frac{1}{\Gamma\left(\zeta\right)}\\
\times\int_{0}^{\infty}d\beta\,\frac{\beta^{\zeta-1}e^{-\beta}}{\mu_{\perp}+\beta/\eta}Z\left(z\sqrt{\mu_{\parallel}+\frac{\beta}{\eta}}\right),\label{eq:RKAP2:DF-Anisotropic-Superstatistics}
\end{multline}
where $Z\left(\psi\right)$ is the well-known Fried \& Conte function
(see Appendix \ref{sec:bMd-bSKd}).
We also need the derivative
$Z_{\kappa}^{\left(\eta,\zeta,\boldsymbol{\mu}\right)\prime}\left(z\right)=dZ_{\kappa}^{\left(\eta,\zeta,\boldsymbol{\mu}\right)}\left(z\right)/dz$,
which is 
\begin{multline}
Z_{\kappa}^{\left(\eta,\zeta,\boldsymbol{\mu}\right)\prime}\left(z\right)=\frac{\eta^{-3/2}}{\mathcal{U}^{\left(\eta,\zeta,\boldsymbol{\mu}\right)}}\frac{1}{\Gamma\left(\zeta\right)}\\
\times\int_{0}^{\infty}d\beta\,\frac{\sqrt{\mu_{\parallel}+\beta/\eta}}{\mu_{\perp}+\beta/\eta}\beta^{\zeta-1}e^{-\beta}Z^{\prime}\left(z\sqrt{\mu_{\parallel}+\frac{\beta}{\eta}}\right).\label{eq:RKAP2:DFP-Anisotropic-Superstatistics}
\end{multline}

While the dispersion functions \eqref{eq:RKAP2:DF-Anisotropic-Superstatistics}
and \eqref{eq:RKAP2:DFP-Anisotropic-Superstatistics} depend solely
on the anisotropy of the regularization parameters,  there
appears, additionally, another dispersion function which depends on the effects of
all anisotropies: on the thermal speed parameters $\left(\theta_{\parallel},\theta_{\perp}\right)$
and on the regularization parameters $\left(\mu_{\parallel},\mu_{\perp}\right)$\@.
Specifically,

\begin{multline}
\mathcal{Z}_{\kappa,\boldsymbol{\theta}}^{\left(\eta,\zeta,\boldsymbol{\mu}\right)}\left(z\right)=\frac{\eta^{-3/2}}{\mathcal{U}^{\left(\eta,\zeta,\boldsymbol{\mu}\right)}}\frac{1}{\Gamma\left(\zeta\right)}\\
\times\int_{0}^{\infty}d\beta\,\frac{\mathscr{A}_{\boldsymbol{\theta}}^{\boldsymbol{\mu}}\left(\beta\right)\beta^{\zeta-1}e^{-\beta}}{\left(\mu_{\perp}+\beta/\eta\right)\left(\mu_{\parallel}+\beta/\eta\right)}Z\left(z\sqrt{\mu_{\parallel}+\frac{\beta}{\eta}}\right),\label{eq:RKAP2:CDF-Anisotropic-Superstatistics}
\end{multline}
which depends on the global anisotropy function 
\begin{equation}
\mathscr{A}_{\boldsymbol{\theta}}^{\boldsymbol{\mu}}\left(\beta\right)=1-\frac{\phi_{\perp}^{2}\left(\beta\right)}{\phi_{\parallel}^{2}\left(\beta\right)}=1-\left(1-A_{\boldsymbol{\theta}}\right)\frac{\mu_{\parallel}+\beta/\eta}{\mu_{\perp}+\beta/\eta},\label{eq:RKAP3:Global_anisotropy_function}
\end{equation}
with $A_{\boldsymbol{\theta}}=1-\theta_{\perp}^{2}/\theta_{\parallel}^{2}$\@.
From the definition of $\mathcal{Z}_{\kappa,\boldsymbol{\theta}}^{\left(\eta,\zeta,\boldsymbol{\mu}\right)}\left(z\right)$, we need to consider two special cases. 
The first is the derivative $\mathcal{Z}_{\kappa,\boldsymbol{\theta}}^{\left(\eta,\zeta,\boldsymbol{\mu}\right)\prime}\left(z\right)=d\mathcal{Z}_{\kappa,\boldsymbol{\theta}}^{\left(\eta,\zeta,\boldsymbol{\mu}\right)}\left(z\right)/dz$,
given by 
\begin{multline}
\mathcal{Z}_{\kappa,\boldsymbol{\theta}}^{\left(\eta,\zeta,\boldsymbol{\mu}\right)\prime}\left(z\right)=\frac{\eta^{-3/2}}{\mathcal{U}^{\left(\eta,\zeta,\boldsymbol{\mu}\right)}}\frac{1}{\Gamma\left(\zeta\right)}\\
\times\int_{0}^{\infty}d\beta\,\frac{\mathscr{A}_{\boldsymbol{\theta}}^{\boldsymbol{\mu}}\left(\beta\right)\beta^{\zeta-1}e^{-\beta}}{\left(\mu_{\perp}+\beta/\eta\right)\sqrt{\mu_{\parallel}+\beta/\eta}}Z^{\prime}\left(z\sqrt{\mu_{\parallel}+\frac{\beta}{\eta}}\right).\label{eq:RKAP2:CDFP-Anisotropic-Superstatistics}
\end{multline}

\begin{subequations}
\label{eq:RKAP2:CDFP-theta_Anisotropic-Superstatistics}

The other relevant case is the $\mu$-isotropic limit of \eqref{eq:RKAP2:CDFP-Anisotropic-Superstatistics},
which reduces to 
\begin{equation}
\mathcal{Z}_{\kappa,\boldsymbol{\theta}}^{\left(\eta,\zeta,\boldsymbol{\mu}\right)\prime}\left(z\right)=A_{\boldsymbol{\theta}}\mathcal{Z}_{\kappa}^{\left(\eta,\zeta,\mu\right)\prime}\left(z\right),
\end{equation}
where 
\begin{multline}
\mathcal{Z}_{\kappa}^{\left(\eta,\zeta,\mu\right)\prime}\left(z\right)=\frac{\eta^{-3/2}\left(\eta\mu\right)^{3/2-\zeta}}{U\left(\zeta,\zeta-\nicefrac{1}{2},\eta\mu\right)}\frac{1}{\Gamma\left(\zeta\right)}\\
\times\int_{0}^{\infty}d\beta\,\frac{\beta^{\zeta-1}e^{-\beta}}{\left(\mu+\beta/\eta\right)^{3/2}}Z^{\prime}\left(z\sqrt{\mu+\frac{\beta}{\eta}}\right).
\end{multline}
\end{subequations}
The representations \eqref{eq:RKAP2:DF-Anisotropic-Superstatistics}
-- \eqref{eq:RKAP2:CDFP-theta_Anisotropic-Superstatistics} of the
new dispersion functions in terms of improper integrals are enough
for their numerical evaluation, as long as efficient routines for
the evaluation of $Z\left(\psi\right)$ and its derivative on the
entire complex plane are available.
Nevertheless, the well-known mathematical properties of the Fried
\& Conte function allows one to derive additional representations
for certain regions of the Argand plane of $\psi$\@. Some of these
properties will be derived below, respectively for expressions \eqref{eq:RKAP2:DF-Anisotropic-Superstatistics},
\eqref{eq:RKAP2:DFP-Anisotropic-Superstatistics}, \eqref{eq:RKAP2:CDFP-Anisotropic-Superstatistics}
and \eqref{eq:RKAP2:CDFP-theta_Anisotropic-Superstatistics}.

\subsubsection{Values at origin}

\begin{subequations}
\label{eq:RKAP3:Values-origin}

Knowing that $Z\left(0\right)=i\sqrt{\pi}$ and $Z^{\prime}(0) = -2$,
by setting $z=0$ in the sequence of dispersion functions and comparing the resulting integrals with the results in section \ref{subsec:Case-z0}, we obtain 
\begin{align}
Z_{\kappa}^{\left(\eta,\zeta,\boldsymbol{\mu}\right)}\left(0\right) & =-i\sqrt{\frac{\pi}{\eta}}\frac{U\left(1,2-\zeta,\eta\mu_{\perp}\right)}{\mathcal{U}^{\left(\eta,\zeta,\boldsymbol{\mu}\right)}}\\
Z_{\kappa}^{\left(\eta,\zeta,\boldsymbol{\mu}\right)\prime}\left(0\right) & =-\frac{2}{\eta}\frac{\mathcal{U}_{-1,0}^{\left(\eta,\zeta,\boldsymbol{\mu}\right)}}{\mathcal{U}^{\left(\eta,\zeta,\boldsymbol{\mu}\right)}}\\
\mathcal{Z}_{\kappa,\boldsymbol{\theta}}^{\left(\eta,\zeta,\boldsymbol{\mu}\right)\prime}\left(0\right) & =-2\left[1-\left(1-A_{\boldsymbol{\theta}}\right)\frac{\mathcal{U}_{-1,1}^{\left(\eta,\zeta,\boldsymbol{\mu}\right)}}{\mathcal{U}^{\left(\eta,\zeta,\boldsymbol{\mu}\right)}}\right]\\
\mathcal{Z}_{\kappa}^{\left(\eta,\zeta,\mu\right)\prime}\left(0\right) & =-2,
\end{align}
where we have also used \eqref{eq:RKAP3:Global_anisotropy_function}.
\end{subequations}

\subsubsection{Power series expansions}

We take the power series expansions for the $Z$ function and its
derivative 
\begin{align*}
Z\left(z\right) & =i\sqrt{\pi}e^{-z^{2}}-2z\sum_{n=0}^{\infty}\frac{\left(-z^{2}\right)^{n}}{\left(\nicefrac{3}{2}\right)_{n}}, \nonumber \\ 
Z^{\prime}\left(z\right) & =-2i\sqrt{\pi}z e^{-z^{2}}-2\sum_{n=0}^{\infty}\frac{\left(-z^{2}\right)^{n}}{\left(\nicefrac{1}{2}\right)_{n}},
\end{align*}
where $\left(a\right)_{n}=\Gamma\left(a+n\right)/\Gamma\left(a\right)$
is the Pochhammer symbol, with $\Gamma\left(z\right)$ being the gamma
function. Inserting these expansions into the representations \eqref{eq:RKAP2:DF-Anisotropic-Superstatistics},
\eqref{eq:RKAP2:DFP-Anisotropic-Superstatistics}, \eqref{eq:RKAP2:CDFP-Anisotropic-Superstatistics}
and \eqref{eq:RKAP2:CDFP-theta_Anisotropic-Superstatistics}, the
ensuing integrals can be readily identified with the results obtained
in Appendix \ref{sec:Definitions-and-properties}, resulting 
\begin{widetext}
\begin{subequations}
\label{eq:Power-series}
\begin{align}
Z_{\kappa}^{\left(\eta,\zeta,\boldsymbol{\mu}\right)}\left(z\right) & =-\frac{2z/\eta}{\mathcal{U}^{\left(\eta,\zeta,\boldsymbol{\mu}\right)}}\sum_{n=0}^{\infty}\frac{\mathcal{U}_{-n-1,0}^{\left(\eta,\zeta,\boldsymbol{\mu}\right)}}{\left(\nicefrac{3}{2}\right)_{n}}\left(-\frac{z^{2}}{\eta}\right)^{n}+iZ_{\kappa,C}^{\left(\eta,\zeta,\boldsymbol{\mu}\right)}\left(z\right)\\
Z_{\kappa}^{\left(\eta,\zeta,\boldsymbol{\mu}\right)\prime}\left(z\right) & =-\frac{2\eta^{-1}}{\mathcal{U}^{\left(\eta,\zeta,\boldsymbol{\mu}\right)}}\sum_{n=0}^{\infty}\frac{\mathcal{U}_{-n-1,0}^{\left(\eta,\zeta,\boldsymbol{\mu}\right)}}{\left(\nicefrac{1}{2}\right)_{n}}\left(-\frac{z^{2}}{\eta}\right)^{n}+iZ_{\kappa,C}^{\left(\eta,\zeta,\boldsymbol{\mu}\right)\prime}\left(z\right)\\
\mathcal{Z}_{\kappa,\boldsymbol{\theta}}^{\left(\eta,\zeta,\boldsymbol{\mu}\right)\prime}\left(z\right) & =-\frac{2}{\mathcal{U}^{\left(\eta,\zeta,\boldsymbol{\mu}\right)}}\sum_{n=0}^{\infty}\frac{\mathcal{V}_{n}^{\left(\eta,\zeta,\boldsymbol{\mu}\right)}}{\left(\nicefrac{1}{2}\right)_{n}}\left(-\frac{z^{2}}{\eta}\right)^{n}+i\mathcal{Z}_{\kappa,\boldsymbol{\theta},C}^{\left(\eta,\zeta,\boldsymbol{\mu}\right)\prime}\left(z\right)\\
\mathcal{Z}_{\kappa}^{\left(\eta,\zeta,\mu\right)\prime}\left(z\right) & =-2\sum_{n=0}^{\infty}\frac{U\left(\zeta,\zeta-\nicefrac{1}{2}+n,\eta\mu\right)}{U\left(\zeta,\zeta-\nicefrac{1}{2},\eta\mu\right)}\frac{\left(-\mu z^{2}\right)^{n}}{\left(\nicefrac{1}{2}\right)_{n}}+i\mathcal{Z}_{\kappa,C}^{\left(\eta,\zeta,\mu\right)\prime}\left(z\right),
\end{align}
where 
\[
\mathcal{V}_{n}^{\left(\eta,\zeta,\boldsymbol{\mu}\right)}=\mathcal{U}_{-n,0}^{\left(\eta,\zeta,\boldsymbol{\mu}\right)}-\left(1-A_{\boldsymbol{\theta}}\right)\mathcal{U}_{-n-1,1}^{\left(\eta,\zeta,\boldsymbol{\mu}\right)}.
\]
\end{subequations}

\begin{subequations}
\label{eq:Continued-functions}

In (\ref{eq:Power-series}a-d), we have also introduced the functions

\begin{align}
Z_{\kappa,C}^{\left(\eta,\zeta,\boldsymbol{\mu}\right)}\left(z\right) & =\sqrt{\pi\mu_{\perp}}\frac{\left(\eta\mu_{\perp}\right)^{\zeta-3/2}}{\mathcal{U}^{\left(\eta,\zeta,\boldsymbol{\mu}\right)}}e^{-\mu_{\parallel}z^{2}}U\left(\zeta,\zeta,\eta\mu_{\perp}\left(1+\frac{z^{2}}{\eta}\right)\right)\\
Z_{\kappa,C}^{\left(\eta,\zeta,\boldsymbol{\mu}\right)\prime}\left(z\right) & =-\frac{2\sqrt{\pi}\eta^{-1}}{\mathcal{U}^{\left(\eta,\zeta,\boldsymbol{\mu}\right)}}\frac{z}{\sqrt{\eta}}e^{-\mu_{\parallel}z^{2}}\left[\left(1+\frac{z^{2}}{\eta}\right)^{-\zeta}-\left(\eta\mu_{\perp}\right)^{\zeta}\mathcal{A}_{\boldsymbol{\mu}}U\left(\zeta,\zeta,\eta\mu_{\perp}\left(1+\frac{z^{2}}{\eta}\right)\right)\right]\\
\mathcal{Z}_{\kappa,\boldsymbol{\theta},C}^{\left(\eta,\zeta,\boldsymbol{\mu}\right)\prime}\left(z\right) & =-2\sqrt{\pi}\frac{\left(\eta\mu_{\perp}\right)^{\zeta-1}}{\mathcal{U}^{\left(\eta,\zeta,\boldsymbol{\mu}\right)}}\frac{z}{\sqrt{\eta}}e^{-\mu_{\parallel}z^{2}}\left[A_{\boldsymbol{\theta}}U\left(\zeta,\zeta,\eta\mu_{\perp}\left(1+\frac{z^{2}}{\eta}\right)\right)+\mathcal{A}_{\boldsymbol{\mu}}\left(1-A_{\boldsymbol{\theta}}\right)U\left(\zeta,\zeta-1,\eta\mu_{\perp}\left(1+\frac{z^{2}}{\eta}\right)\right)\right]\\
\mathcal{Z}_{\kappa,C}^{\left(\eta,\zeta,\mu\right)\prime}\left(z\right) & =-2\sqrt{\pi\mu}ze^{-\mu z^{2}}\frac{U\left(\zeta,\zeta,\eta\mu\left(1+z^{2}/\eta\right)\right)}{U\left(\zeta,\zeta-\nicefrac{1}{2},\eta\mu\right)}.
\end{align}
\end{subequations}

\subsubsection{Asymptotic expansions}

Similarly, we start from the known asymptotic expansions 
\[
\begin{aligned}Z\left(z\right) & \sim-\frac{1}{z}\sum_{k=0}\frac{\left(\nicefrac{1}{2}\right)_{k}}{z^{2k}}+\sigma i\sqrt{\pi}e^{-z^{2}}, & Z^{\prime}\left(z\right) & \sim\sum_{k=0}\frac{\left(\nicefrac{3}{2}\right)_{k}}{z^{2\left(k+1\right)}}-2\sigma i\sqrt{\pi}z e^{-z^{2}},\end{aligned}
\text{where }\sigma=\begin{cases}
0, & \Im z > 0\\
1, & \Im z = 0\\
2, & \Im z < 0.
\end{cases}
\]

\begin{subequations}
\label{eq:Asymptotic-expansions}

Proceeding in a similar manner as in the previous sections, we obtain
the following asymptotic expansions 
\begin{align*}
Z_{\kappa}^{\left(\eta,\zeta,\boldsymbol{\mu}\right)}\left(z\right) & \sim-\frac{1}{z}\sum_{k=0}\left(\frac{1}{2}\right)_{k}\frac{\mathcal{U}_{k,0}^{\left(\eta,\zeta,\boldsymbol{\mu}\right)}}{\mathcal{U}^{\left(\eta,\zeta,\boldsymbol{\mu}\right)}}\left(\frac{\eta}{z^{2}}\right)^{k}+\sigma iZ_{\kappa,C}^{\left(\eta,\zeta,\boldsymbol{\mu}\right)}\left(z\right)\\
Z_{\kappa}^{\left(\eta,\zeta,\boldsymbol{\mu}\right)\prime}\left(z\right) & \sim\frac{1}{z^{2}}\sum_{k=0}\left(\frac{3}{2}\right)_{k}\frac{\mathcal{U}_{k,0}^{\left(\eta,\zeta,\boldsymbol{\mu}\right)}}{\mathcal{U}^{\left(\eta,\zeta,\boldsymbol{\mu}\right)}}\left(\frac{\eta}{z^{2}}\right)^{k}+\sigma iZ_{\kappa,C}^{\left(\eta,\zeta,\boldsymbol{\mu}\right)\prime}\left(z\right)\\
\mathcal{Z}_{\kappa,\boldsymbol{\theta}}^{\left(\eta,\zeta,\boldsymbol{\mu}\right)\prime}\left(z\right) & \sim\frac{\eta}{z^{2}}\sum_{k=0}\left(\frac{3}{2}\right)_{k}\left[\frac{\mathcal{U}_{k+1,0}^{\left(\eta,\zeta,\boldsymbol{\mu}\right)}}{\mathcal{U}^{\left(\eta,\zeta,\boldsymbol{\mu}\right)}}-\left(1-A_{\boldsymbol{\theta}}\right)\frac{\mathcal{U}_{k,1}^{\left(\eta,\zeta,\boldsymbol{\mu}\right)}}{\mathcal{U}^{\left(\eta,\zeta,\boldsymbol{\mu}\right)}}\right]\left(\frac{\eta}{z^{2}}\right)^{k}+\sigma i\mathcal{Z}_{\kappa,\boldsymbol{\theta},C}^{\left(\eta,\zeta,\boldsymbol{\mu}\right)\prime}\left(z\right)\\
\mathcal{Z}_{\kappa}^{\left(\eta,\zeta,\mu\right)\prime}\left(z\right) & \sim\frac{\eta}{z^{2}}\sum_{k=0}\left(\frac{3}{2}\right)_{k}\frac{U\left(\nicefrac{5}{2}+k,\nicefrac{7}{2}+k-\zeta,\mu\eta\right)}{U\left(\nicefrac{3}{2},\nicefrac{5}{2}-\zeta,\eta\mu\right)}\left(\frac{\eta}{z^{2}}\right)^{k}+\sigma i\mathcal{Z}_{\kappa,C}^{\left(\eta,\zeta,\mu\right)\prime}\left(z\right),
\end{align*}
valid for $\left|z\right|\gg\sqrt{\eta}$, also in terms of the functions
(\ref{eq:Continued-functions}a-d).
\end{subequations}
\end{widetext}

\section{Instabilities from semi-anisotropic distributions}
\label{sec:Semi-anisotropic}

The expressions derived in Sec. \ref{sec:Theoretical-formulation}
will now be applied to the study of wave instabilities arising due
to nonthermal features in the electron VDF,
such as temperature anisotropies and drift speeds. We examine the instabilities of electromagnetic, circularly-polarized modes given by the solutions of the dispersion equation (\ref{eq:RKAP2:DE-RL}).

We assume a plasma of electrons and ions (mostly protons) typical of the solar wind, often described with a three-component electron population: at low energies the core described by a dbM VDF, a more tenuous suprathermal halo, a beam or strahl component. For simplicity - and supported by recent observations in the young solar wind \citep{Halekas+22/09} - here we neglect the halo component, adopting instead a two-component electron VDF comprising the core (subscript $c$) plus the strahl (subscript $s$):
\begin{equation}
f_{e}\left(v_{\parallel},v_{\perp}\right)=\frac{n_{c}}{n}f_{c}\left(v_{\parallel},v_{\perp}\right)+\frac{n_{s}}{n}f_{s}\left(v_{\parallel},v_{\perp}\right).\label{eq:Core-Strahl_model}
\end{equation}
 The relative densities and drift speeds satisfy the usual constraints
of zero net charge $n_{i}=n_{e}=n_{c}+n_{s}$ and zero current density $n_{c}U_{\parallel c}+n_{s}U_{\parallel s}=0$.
The dispersion equation will be solved in the ion's rest frame $\left(U_{\parallel i}=0\right)$.

The components of the susceptibility tensor (\ref{eq:RKAP2:DT-magnetized-parallel}a-c)
will be analytically evaluated for the core-strahl model (\ref{eq:Core-Strahl_model})\@.
The strahl population will be modelled by a drifting bi-RKD (\ref{eq:RKAP3:RKD-anisotropic-beam-generalized}), identified by the acronym dbRK\@. 
For this population, the partial susceptibilities are given by Eqs. (\ref{eq:DT-bRKDd}a, b), evaluated with the new dispersion functions (\ref{eq:RKAP2:DF-Anisotropic-Superstatistics}) and (\ref{eq:RKAP2:CDFP-Anisotropic-Superstatistics}).
%
The susceptibility tensor components for the dbM core can be found in the literature and are
listed in Appendix \ref{sec:bMd-bSKd}\@. 
%
Since we are interested in the instabilities driven by free energy sources contained in the electronic component of the plasma, the ion contribution is not relevant for this study, and so they will be modelled by an isotropic Maxwellian distribution (see   \ref{eq:Maxwellian-VDF}) with $U_\parallel = 0$ and $\theta_\parallel=\theta_\perp$.

Once the total susceptibility tensor of the core-strahl system is evaluated, the dispersion equation (\ref{eq:RKAP2:DE-RL}) is numerically solved in order to provide the complex function 
\[
\omega\left(k_{\parallel}\right)=\omega_{r}\left(k_{\parallel}\right)+i\gamma\left(k_{\parallel}\right),
\]
with the wave frequency dispersion relation $\omega_{r}\left(k_{\parallel}\right)$
for the $R$ and $L$ circularly-polarized modes, and the associated damping/growth rate as $\gamma\left(k_{\parallel}\right)$\@. 
This task is accomplished by a specific computer code developed in Modern Fortran that numerically solves Eq. (\ref{eq:RKAP2:DE-RL}) in any frequency range and for an arbitrary number and combinations of plasma populations modelled by Maxwellian (dbM), standard Kappa (dbSK) or regularized Kappa (dbRK) distributions.
This approach is called here \emph{semi-analytic} and this is the first time it is applied to analyse kinetic instabilities excited by anisotropic RKDs.

The semi-analytic approach will be independently complemented and tested by another approach, employing the Arbitrary Linear Plasma Solver (ALPS),\cite{Verscharen_2018}
which is a parallelized numerical solver, developed in Modern Fortran, that computes solutions to the most general dispersion relation under a wide range of plasma conditions, including both hot non-relativistic and relativistic regimes. Its flexibility enables the treatment of an arbitrary number of particle species with equilibrium distribution functions~$f_{0j}$, and it allows for any propagation direction relative to the unperturbed background magnetic field.
To run ALPS, one merely needs to provide numerical input for the equilibrium distributions~$f_{0,j}(p_{\perp},p_{\|})$, where~$p_{\|}$ and~$p_{\perp}$ represent the momentum components parallel and perpendicular to the background field. 
Moreover, initial guesses for the real $\omega_r$ and imaginary $\gamma$, must be specified.
The code applies an iterative Newton–secant scheme, yielding the dispersion relation in terms of $\omega_r(\boldsymbol{k})$ and $\gamma(\boldsymbol{k})$. Further details on the implementation and functionality of ALPS and its applicability can be found in Refs~\onlinecite{Verscharen_2018,Schroeder+25/03,Schroeder+25/07,Schroeder+26/04}. Note, in particular, that ALPS is not informed about any specific form of the VDFs and, therefore, its 'fully numerical' results are excellent tests for those obtained with the semi-analytical approach. 

The excitation of wave instabilities will be first studied for a strahl population modelled by a dbRK VDF with temperature anisotropy $\left(T_{\perp s}\neq T_{\parallel s}\right)$, but with isotropic regularization $\left(\mu_{\perp s}=\mu_{\parallel s}=\mu\right)$, a situation already analyzed by Ref. \onlinecite{Schroeder+25/07}, but employing solely a purely numerical approach with the ALPS code.
This initial comparison between the semi-analytic and the fully numerical approaches is worthy by itself because the limit of isotropic $\mu$ is an important particular case of the formalism developed Sec. \ref{sec:Theoretical-formulation} and the results presented here will serve as a mutual validation of both approaches.
In all cases, the parameters in the generalized expression (\ref{eq:RKAP3:RKD-anisotropic-beam-generalized}) were fixed to $\eta =\kappa$, and $\zeta =\kappa+1$.
%
%
Each population is described by the following nondimensional quantities 
\begin{align*}
\beta_{a} & =\frac{4\pi n_{a} m_{a} \theta_{\parallel a}^{2}}{B_{0}^{2}}, 
& u_{s} & =\frac{U_{\parallel s}}{c_l}, & \eta & =\frac{n_{s}}{n_{e}}, & r_{Ta} & =\frac{T_{\perp a} }{T_{\| a} },
\end{align*}
%
%
where $\beta_{a}$ is 
related to
the beta parameter for population $a$ ($a=c,s$), $B_{0}$ is the magnetic field and $c_l$ is the speed of light in vacuum. 
With this choice of nondimensional parameters, $\beta_a$ is not the physical beta for the plasma population, because $\theta_{\parallel a}$ is a constant parameter in our formulation.  The actual measure of the second moment of the VDF is Eq. (\ref{eq:Temperatures_bSKd}) for the standard Kappa, or Eq. (\ref{eq:RKAP3:bRKd-Kinetic_temperatures}) for the regularized Kappa distribution.
Hereafter, we adopt $\eta$ for the density ratio, since the homonymous parameter in Eq.~(\ref{eq:RKAP3:RKD-anisotropic-beam-generalized}) is already set equal to $\kappa$\@.
It is also worth pointing out that in the present case, according to (\ref{eq:RKAP3:bRKd-Kinetic_temperatures}) and (\ref{eq:CalU-Isotropic_mu}), the temperature ratio $r_{Ta}$ is independent of the regularization parameter, similarly to what happens with a standard bi-Kappa distribution.



\begin{figure*}[!t]
\includegraphics[width=1\textwidth]{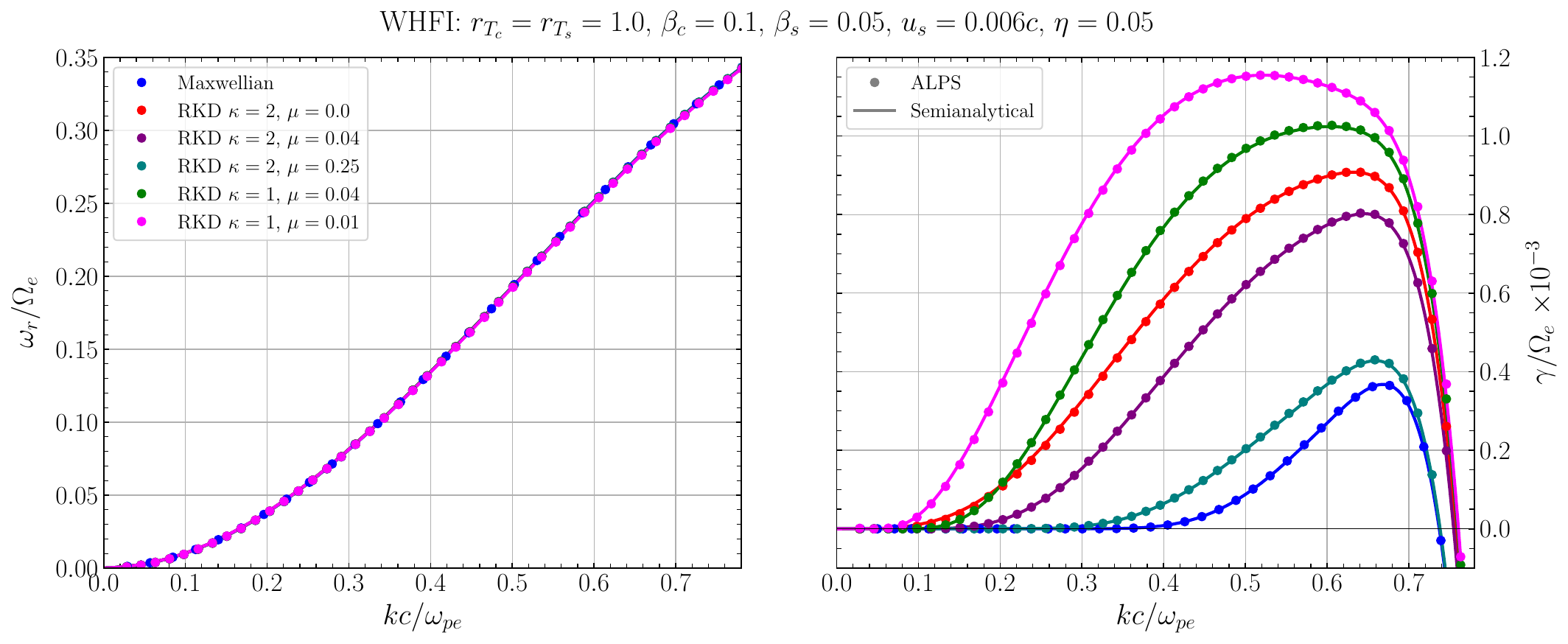}
\caption{Dispersion relations (left panel) and growth rates (right panel) of
WHFI triggered by strahl populations with different values of the
$\kappa$ and the regularization parameter. Continuous lines
were produced by the semi-analytical approach, whereas the dots are
results obtained with ALPS\@. The same conventions and colors are
adopted in the following figures.}
\label{fig:disp_WHFI}
\end{figure*}

Instabilities generated by the electron strahl, which is very often the main heat-flux carrier in the solar wind, are known as heat-flux instabilities (HFIs). 
Here these unstable solutions will also accumulate the effects of the second source of free energy that is the temperature anisotropy.
Figure \ref{fig:disp_WHFI},
which reproduces Fig. 6 of Ref.~\onlinecite{Schroeder+25/07},
shows the results for the ``pure'' whistler heat-flux instability (WHFI), occurring in the $R$ mode for isotropic temperature,  $(r_{Ti} = r_{Tc} = r_{Ts} = 1 )$. 
The sole free energy source resides in the relative core-strahl drift.
Results obtained from the semi-analytical approach are depicted as continuous lines, whereas the ALPS results are identified by the dots. 
The left panel shows the frequency dispersion relations for each parameter set (indicated in the legend and specifically selected to trigger the instability), whereas the right panel shows the associated growth rates.
In particular, notice that WHFI is excited by relatively slow beams $\left(u_{s} = 0.006\right)$\@.
Despite the apparently small value of the beam speed, the parameters chosen for Fig. 1  reproduce the conditions close to the maximum growth rate for a system with the same values of $\eta$, $\beta_c$ and $\omega_{pe}/|\Omega_e| = 100$.\cite{{Shaaban+18/10}}

\begin{figure*}[!t]
\includegraphics[width=1\textwidth]{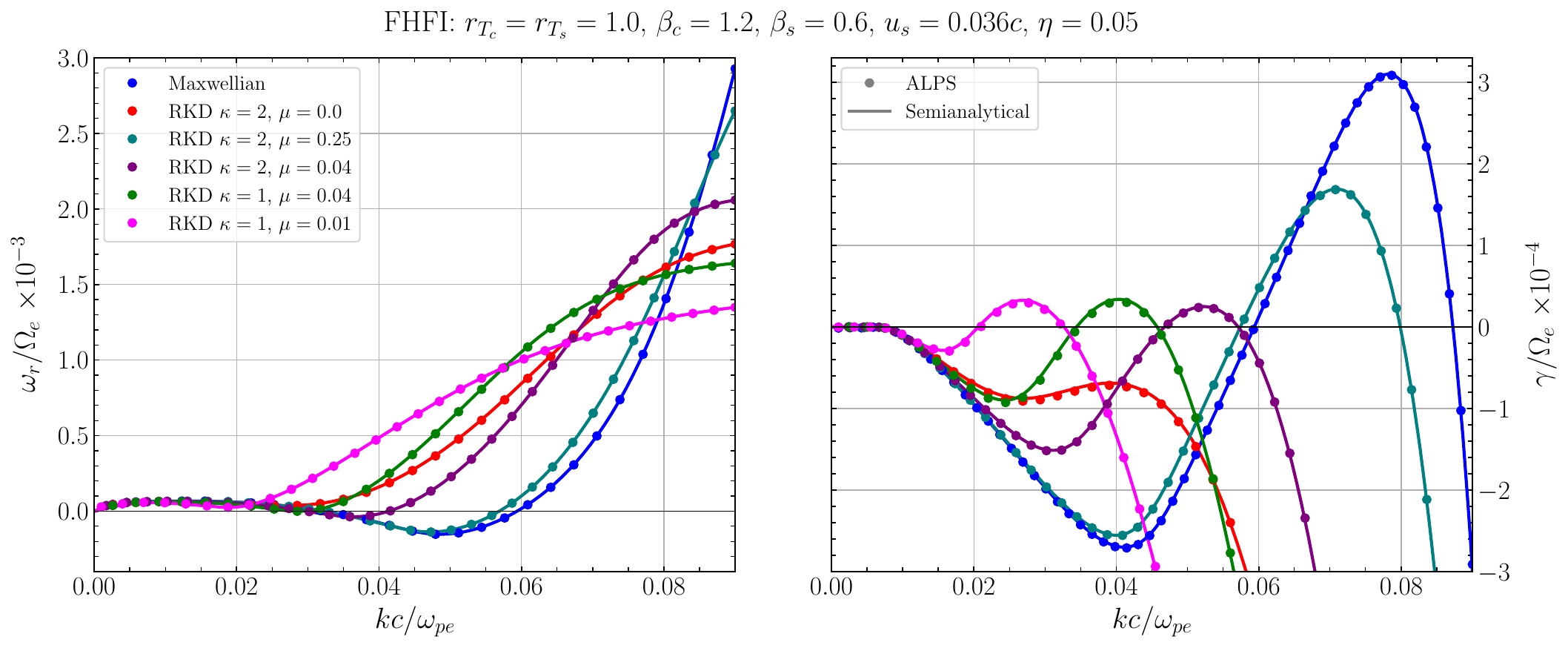}
\caption{FHFI in the $L$ mode excited by a streaming strahl. The beam is still
isotropic around the drift and the other parameters are set in order
to excite the instability.}
\label{fig:disp_FHFI}
\end{figure*}

The unstable solutions were obtained for several combinations of thermal or
suprathermal parameters. In the first run (blue line and dots), the
strahl component was modelled by a dbM VDF.
In the semi-analytical approach, the corresponding $\chi_{1,2}^{\mathrm{\rm dbM}}\left(\omega,k_{\parallel}\right)$
tensor components were evaluated by Eqs. (\ref{eq:ST-bMd}a, b)\@.
One can observe that the instability is already present in the Maxwellian
case.
The red lines and dots depict the results from a strahl described by a drifting bi-Kappa (dbK) VDF with $\kappa=2$,
close to the allowed minimum value of 1.5 for which the second moment can still be evaluated (see Eq.~\ref{eq:Temperatures_bSKd})\@.
The ALPS code reproduces this situation by setting the regularization parameter $\mu=0$ when the VDF grid is generated. 
For the semi-analytical approach, the dbK VDF  (\ref{eq:KAP6:KappaDistFunction}) was adopted.
In this case, the $\chi_{1,2}^{\mathrm{(dbK)}} \left(\omega,k_{\parallel}\right)$
tensor components are given by Eqs. (\ref{eq:ST-bSKd}a, b)\@. 
One can observe that the growth rate for the instability excited by a suprathermal plasma is always higher than that excited by a Maxwellian strahl.

The remaining lines and dots plotted in Fig. 1 show the WHFI excited by RKD strahls, and now the semi-analytical approach employs the new formulas developed here for the evaluation of the  $\chi_{1,2}^{\mathrm{(dbRK)}} \left(\omega,k_{\parallel}\right)$
components given by Eqs. (\ref{eq:DT-bRKDd}a, b)\@. 
The purple line and dots show the results for a dbRK VDF with the same value of $\kappa=2$, but with a finite $\mu=0.04$\@.
Notice that the instability is somewhat smouldered as compared with the SKD result, due to the effect of the (relatively large) regularization parameter. 
The same trend can be observed in the cyan results, again obtained with $\kappa=2$ but with a larger $\mu=0.25$.
In this case, the instability is even more suppressed due to a less suprathermal nature of the RKD, within a lower energy range above which the VDF behaves basically as Maxwellian.

The opposite behavior is observed if the suprathermal feature of the strahl is increased. 
The green results are obtained for a lower $\kappa=1$ and $\mu=0.04$, and now the VDF  retains its suprathermality for larger energies, leading to enhanced growth rate, as compared to the pure dbK case. 
%
For a lower energy cut-off (at even higher levels), the strahl is able to generate stronger instabilities, as exemplified by the magenta results, where $\kappa=1$ and $\mu=0.01$\@. 
In summary, the higher the suprathermality level in the strahl component, the higher the growth rates of the WHFI.
On the other hand, the $R$-mode frequency in the left panel of Fig.~\ref{fig:disp_WHFI} is completely unaffected by the thermal nature of the beam.
The final worthwhile observation  
is the excellent agreement between
the semi-analytical approach and the purely numerical results obtained from ALPS\@. 
This remarkable agreement serves to validate the results obtained from both approaches.

Figure \ref{fig:disp_FHFI} shows results obtained from a set of parameters able to generate the firehose heat-flux instability (FHFI), which occurs in the $L$ mode. 
The FHFI needs relatively higher values of beta parameters and higher drift speeds than the WHFI\@. 
As above, 
the different results were generated by varying the degree of suprathermality of the strahl.
The first noticeable feature is once more the excellent agreement between the results obtained from both approaches adopted in this work. 

Differently from Fig. \ref{fig:disp_WHFI}, though, we now observe that both the real $\left(\omega_{r}\right)$ and imaginary $\left(\gamma\right)$ parts of the solutions are greatly affected by the strahl's thermality level. 
The Maxwellian result (blue curve and dots) shows that the wave can exhibit growth at a relatively high spectral range $(0.06 \lesssim kc/\omega_{pe} \lesssim 0.088)$, while at the same time exhibits a region of negative phase velocity in the range $0.035 \lesssim kc / \omega_{pe} \lesssim 0.06$\@.
However, a completely different picture emerges with a suprathermal strahl. 
The pure dbK case (red) with $\kappa=2$ shows that the instability is completely suppressed while at the same time the wave is forward-propagating throughout the spectral range. 
The Maxwellian result can then be recovered from a RKD strahl when the regularization parameter is high, as can be observed in the purple ($\kappa=2$, $\mu=0.04$) and cyan ($\kappa=2$, $\mu=0.25$) results, with which the thermal case can be asymptotically reproduced.

When the suprathermality of the strahl increases, a somewhat different picture emerges. 
The green ($\kappa=1$, $\mu=0.04$) and magenta
($\kappa=1$, $\mu=0.01$) results show that as the suprathermality
level increases, the FHFI occurs in increasingly lower spectral ranges. 
The figure shows quite different results between the dbK and the dbRK cases, and these results suggest that the maximum growth rates and spectral ranges of the FHFI can depend quite differently on the level of suprathermality of the strahl.

Since the electrons observed in the solar wind also show conspicuous anisotropies in the VDF, it is worthwhile to add the effect of temperature anisotropy to the heat
flux carried by the strahl. 
In this case the strahl will provide two free energy sources capable of triggering instabilities. 
A thorough analysis of the complex interplay between the effects of both energy sources was performed by Refs.~\onlinecite{Shaaban+18/10} and \onlinecite{Shaaban+18/08} for the case when the strahl is modelled by a SKD\@. 
In the present work, we will only present some representative cases, but with a RKD strahl, reproducing the cases discussed by Ref.~\onlinecite{Schroeder+25/07}, which were obtained with the purely numerical approach.

\begin{figure*}[ht!]
\includegraphics[width=1\textwidth]{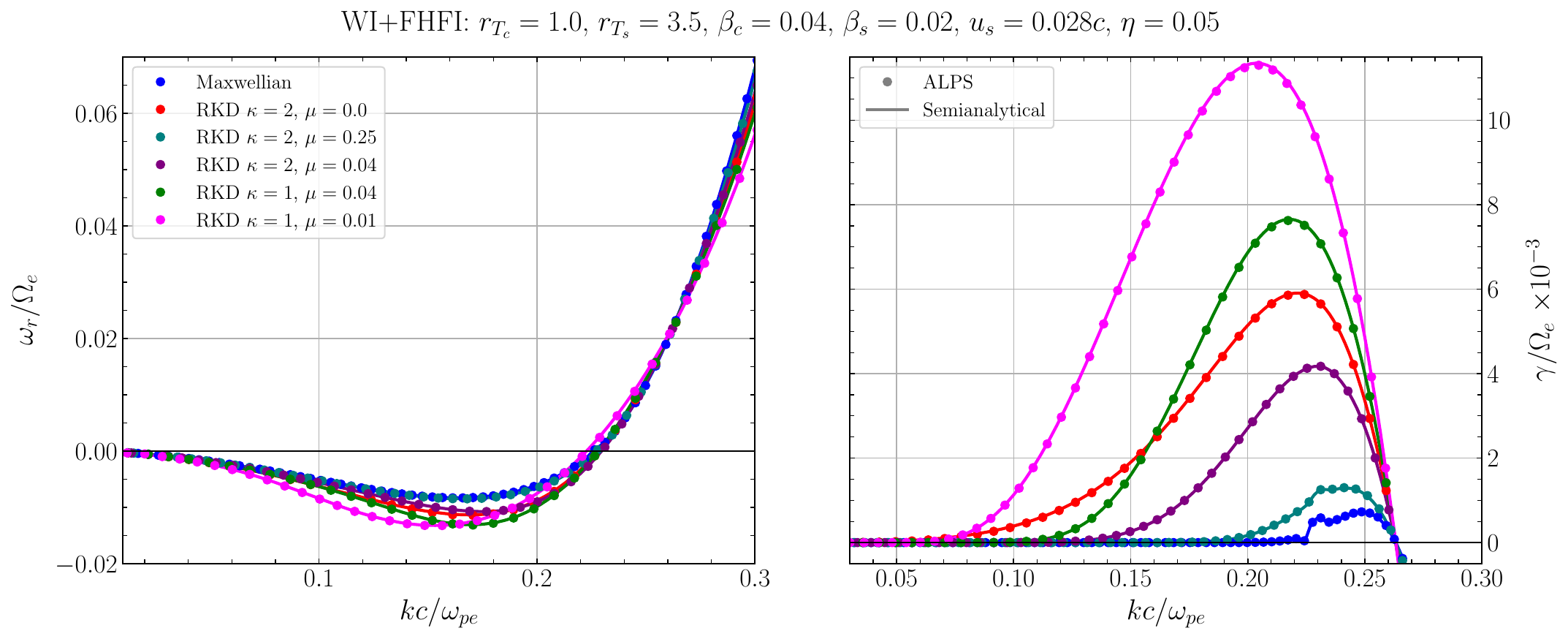}
\caption{Instability occurring in the $L$ mode due to the combined effect
of large heat flux and $T_{\perp s}>T_{\parallel s}$ anisotropy.}
\label{fig:disp_WI_FHFI}
\end{figure*}

The first case considers a beam with large perpendicular temperature
$\left(T_{\perp s}>T_{\parallel s}\right)$, such as is
the case of Fig. \ref{fig:disp_WI_FHFI}\@. Without drift $\left(u_{s}=0\right)$,
the electrons would typically excite the whistler instability acting
on the right-handed $\left(R\right)$ circularly-polarized mode.\cite{Gary05}
However, in the presence of a strong beam $\left(u_{s}=0.028\right)$,
the combined effect destabilizes the $L$ mode.

While the dispersion relation (left panel) is weakly affected by changes
of the RKD parameters, the right panel of Fig. \ref{fig:disp_WI_FHFI}
shows that the growth rate can be greatly enhanced. There are already
substantial increases in the maximum growth rate and in the breadth
of the unstable spectral range due to the transition from Maxwellian
(blue) to SKD (red), as in the previous examples. As in the previous
cases, a RKD strahl can tend asymptotically back to the Maxwellian
result if one increases the regularization parameter (purple and cyan
results).

The opposite effect is once again obtained when one decreases the
kappa parameter (to $\kappa=1$)\@. The green $\left(\mu=0.04\right)$
and magenta $\left(\mu=0.01\right)$ plots in Fig. \ref{fig:disp_WI_FHFI}
show how extreme suprathermal strahls can fuel the instability.

\begin{figure*}[ht!]
\includegraphics[width=1\textwidth]{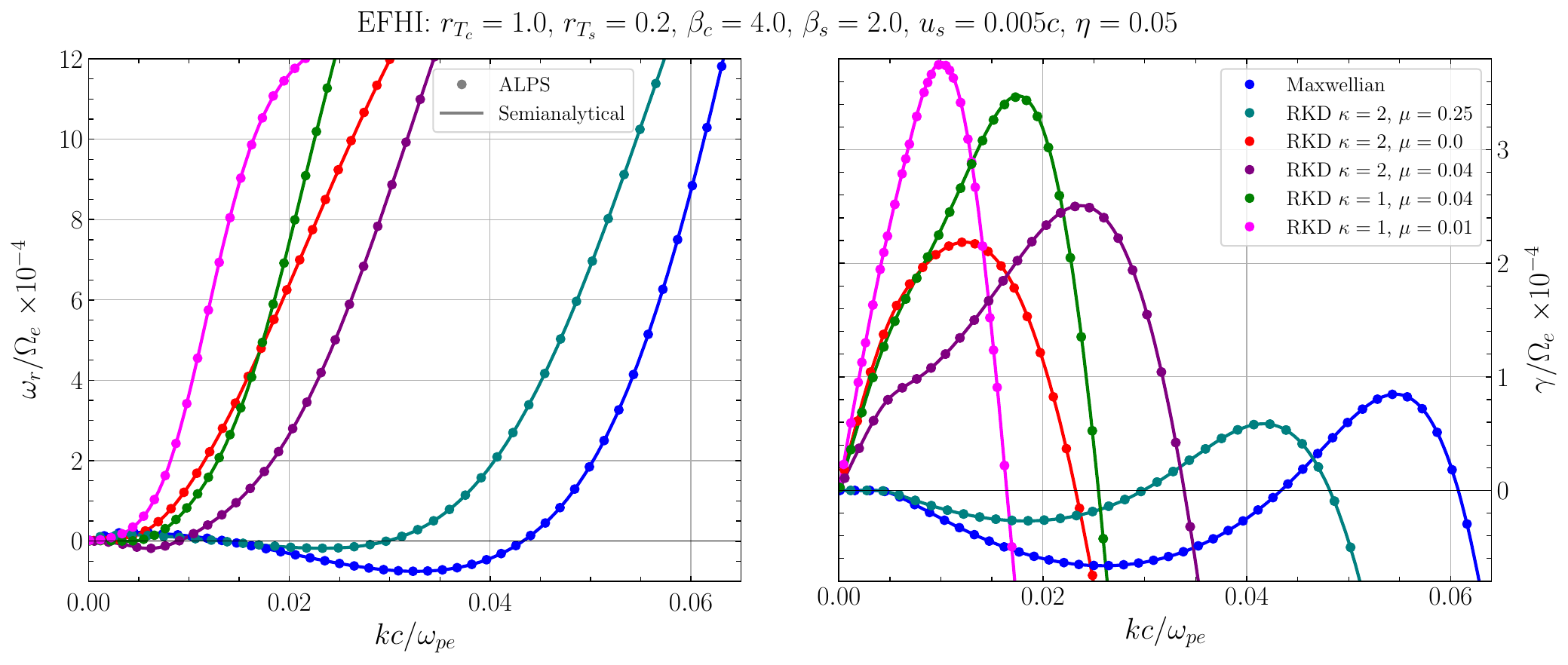}
\caption{Instability occurring in the $L$ mode due to the combined effect
of moderate heat flux and $T_{\perp s}<T_{\parallel s}$
anisotropy.}
\label{fig:disp_EFHI}
\end{figure*}

As a final result in this section, Fig. \ref{fig:disp_EFHI} shows
how inverting the temperature anisotropy to $T_{\perp s}<T_{\parallel s}$
in combination with a moderate heat flux can still excite the $L$
mode.

Without the drift, the anisotropy $r_{Ts}<1$ would equally excite
the electron firehose instability in the $L$ mode. The presence of
a moderate beam speed can exacerbate the effect and changes on the
suprathermality level of the strahl (beam) provoke even further modifications,
this time in both $\omega_{r}$ and $\gamma$.

A great leap is observed in the transition $\text{Maxwellian (blue)}\longrightarrow \text{SKD (red)}$\@. The dispersion relation of the latter case displays
a steep slope in the short wavenumber region, while the growth rate
also increases and moves to a lower spectral range. With a RKD beam,
the results are roughly close to the SKD case, but can once more tend
asymptotically back to the thermal situation. The only noticeable exception
is shown by the purple curve, when the growth rate actually increases,
before reducing back to the Maxwellian result.

\section{Instabilities from fully anisotropic distributions} \label{sec:Fully-anisotropic}

The cases analysed in Sec. \ref{sec:Semi-anisotropic} were restricted
to the situation where the regularization parameter of the regularized
distribution (\ref{eq:RKAP3:RKD-anisotropic-beam-generalized}) is
the same in both directions (\emph{i. e.}, $\mu_{\perp}=\mu_{\parallel}=\mu$)\@.
In this section, this limitation will be relaxed and the effects of
anisotropies both
in the thermal velocity spread parameters $(\theta_\perp \neq \theta_\parallel)$
and in the regularization parameter $\left(\mu_{\perp}\neq\mu_{\parallel}\right)$
will be considered.

As it was already pointed out, in the fully-anisotropic situation, the temperature ratio depends not only the thermal velocity spread parameters $\theta_{\perp(\parallel)}$, but also on the regularization parameters $\mu_{\perp(\parallel)}$\@.  According to (\ref{eq:RKAP3:bRKd-Kinetic_temperatures}), 
\begin{displaymath}
r_{T} = 
\frac{T^{(\eta,\zeta,\boldsymbol{\mu})}_{\perp,\kappa,\boldsymbol{\theta}}}{T^{(\eta,\zeta,\boldsymbol{\mu})}_{\parallel,\kappa,\boldsymbol{\theta}}}=\frac{\theta^{2}_{\perp}}{\theta^{2}_{\parallel}}\frac{\mathcal{U}^{\left(\eta,\zeta,\boldsymbol{\mu}\right)}_{0,1}}{\mathcal{U}^{\left(\eta,\zeta,\boldsymbol{\mu}\right)}_{1,0}}.
\end{displaymath}
Hence, in the following figures, we will provide for the strahl component, instead of $r_{Ts}$ (which was used for Maxwellian core and halo/strahl with  semi-isotropic RKDs ), the ratio
\begin{displaymath}
r_{\theta s} = \frac{\theta_{\perp s}^2}{\theta_{\parallel s}^2}
\end{displaymath}
and then evaluate the temperature ratio $r_{Ts}$ from the formula above.

From a purely argumentative point of view, it stands to reason that
the same forcing mechanisms responsible for the deformation of an
otherwise perfectly isotropic VDF either on the parallel or on the
perpendicular directions relative to the ambient magnetic field should
act on the regularization parameter as well. 
The relevance of the choice of an anisotropic cut-off is not so much practical but rather closer to reality.
The existence of such cut-offs can be justified by the fact that,
due to the finite time and spatial scales of a given physical system,
any heating or acceleration process has limited effects, thereby constraining the energy ranges or levels of the plasma populations in question. 
Since these processes 
occur in different scales, they
are in general different along and perpendicular to the local magnetic background field, 
and consequently
the corresponding energy limits determining the cutoffs of the VDFs are also expected to be unequal. 
While not modelled in the present study, even the form of the cut-off, here chosen to be exponential, may be different, like, e.g., in the case of a truncation of a distribution to particle escape. 
Therefore, the assumption of an isotropic cutoff appears to be a rather special choice.
Although the use of two distinct regularization parameters, $\mu_{\perp}$ and $\mu_{\parallel}$, complicates both the fitting and the interpretation of in-situ measured distributions, this complexity, as shown above, allows for a more realistic treatment of these distributions and their implications.

In this work, the effects of anisotropic $\mu$s on the heat flux
instabilities are analysed for the first time, again with both semi-analytical
and numerical approaches. The former is readily accomplished with
the formul\ae\ 
developed in Sec. \ref{sec:Theoretical-formulation}
and Appendix \ref{sec:Definitions-and-properties}, whilst the latter
is realized by 
setting $\mu_{\|} \neq \mu_{\perp}$ when the VDF grid is created with the ALPS code.%


In this first treatment, we will take the most extreme cases of suprathermal
distributions considered in Figs. \ref{fig:disp_WHFI} -- \ref{fig:disp_EFHI},
namely, cases with $\kappa=1$, and vary the values of $\mu_{\perp}$
and $\mu_{\parallel}$ around the reference value $\mu=0.01$\@.

\begin{figure*}[ht!]
\includegraphics[width=1\textwidth]{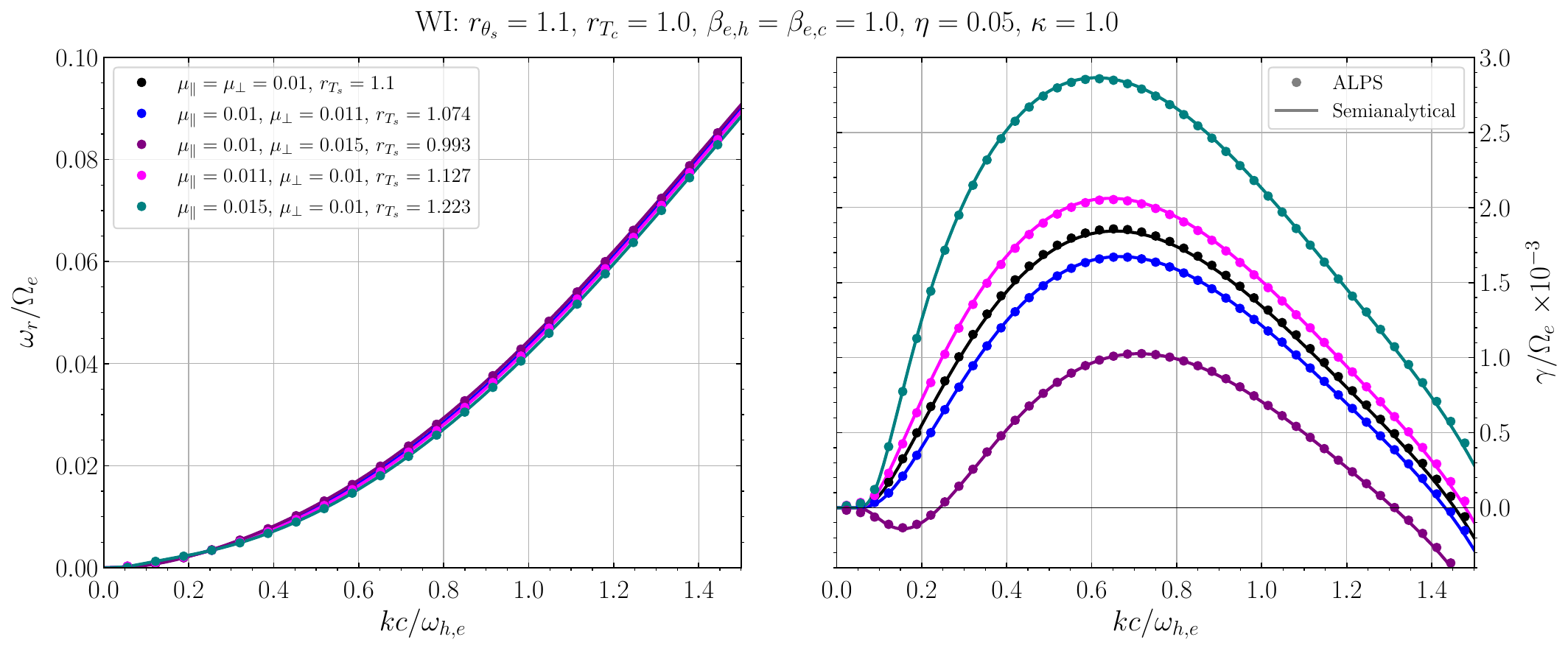}
\caption{Whistler instability generated by anisotropy $r_{\theta_{s}}>1$
in the strahl (or halo) when $\kappa=1$, around the reference value
$\mu_{\perp s}=\mu_{\parallel s}=0.01$.}
\label{fig:disp_WI_cutoff}
\end{figure*}

Figure \ref{fig:disp_WI_cutoff} shows a case that reproduces the
conditions that excite the whistler instability (WI) in the $R$ mode
only due to the anisotropy $\theta_{\perp s} > \theta_{\parallel s}$,
without the supplemental free energy provided by a beam. This parameter
set can reproduce the conditions in a core + halo system.

Since the dispersion relation is largely unaffected by the different
cases, we will concentrate on the growth rate. The reference case
$\mu=0.01$ is depicted by the black curve and dots. The WI is triggered
by the larger dispersion of electron velocities in the perpendicular
direction than in the parallel direction. Thus, an increase of $\mu_{\perp}$
compared to $\mu_{\parallel}$ will move the distribution's cut-off
in the perpendicular direction to lower $v_{\perp}$, meaning that
the degree of suprathermality of the distribution in that direction
is reduced and it tends ever more to behave as a Maxwellian there,
while retaining the original suprathermal characteristic in the parallel
direction. Since a suprathermal plasma tends to support higher growth
rates for the WI than a Maxwellian, we can in this way understand
the effect depicted by the blue $\left(\mu_{\perp}=0.011\right)$
and purple $\left(\mu_{\perp}=0.015\right)$ curves in Fig. \ref{fig:disp_WI_cutoff},
which show a reduction in growth rate due to the incease of $\mu_{\perp}$,
keeping $\mu_{\parallel}=0.01$.

An alternative interpretation of these results is provided by the
fact that an increase in $\mu_{\perp}$ will effectively reduce the
temperature anisotropy, thereby leading to lower growth rates for
the whistler instability.
This interpretation is corroborated by the depicted values of the temperature ratios: an increase of $\mu_\perp$ results in a reduction of $r_{Ts}$, leading to smaller growth ratios.

The opposite effect is seen when $\mu_{\perp}=0.01$ is fixed and
$\mu_{\parallel}$ is increased, as is the case of the magenta $\left(\mu_{\parallel}=0.011\right)$
and cyan $\left(\mu_{\parallel}=0.015\right)$ results. The instability
grows with $\mu_{\parallel}$ and this behavior can be understood
because in this case the effective anisotropy responsible for the
WI is also increased
(the temperature ratios increase).

\begin{figure*}[ht!]
\includegraphics[width=1\textwidth]{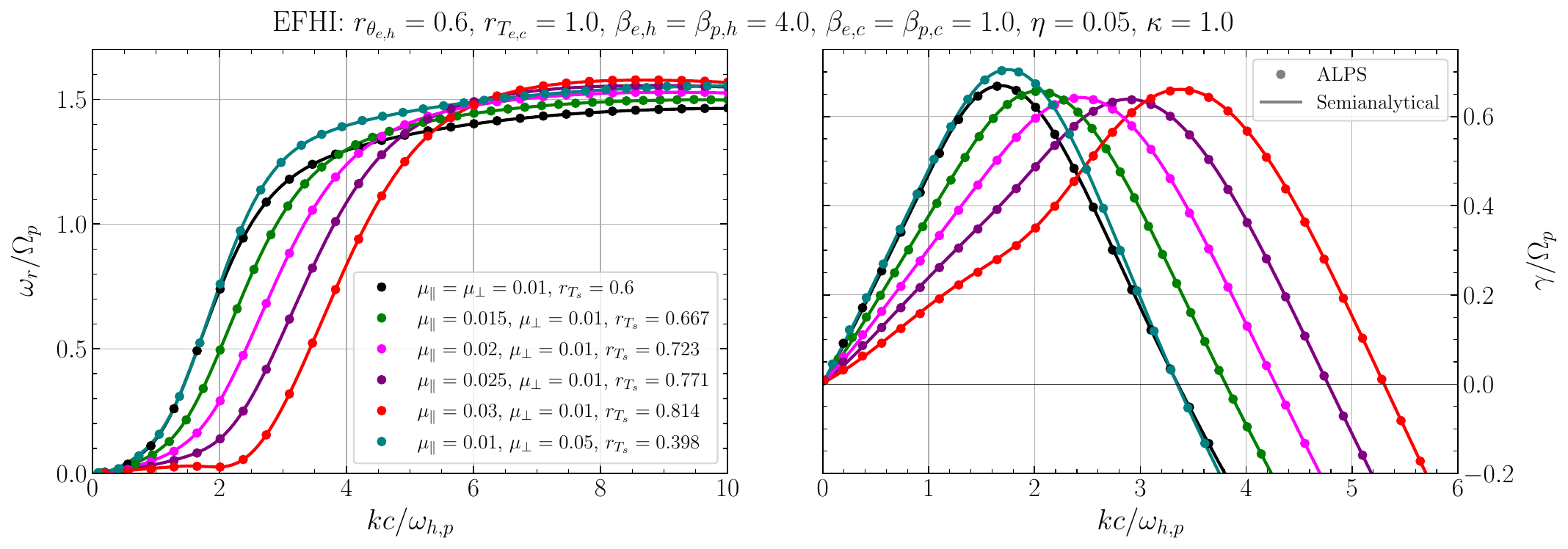}
\caption{Electron firehose instability generated by anisotropy $r_{\theta_{s}}<1$ in the electron halo
when $\kappa=1$, around the reference value $\mu_{\perp h}=\mu_{\parallel h}=0.01$.
Here, a core (Maxwellian) + halo (RKD, $\kappa_{i}=1$, $\mu_{i}=0.01$)
model was also adopted for the ions. All populations are isotropic,
except for the electron halo.}
\label{fig:disp_EFHI_cutoff}
\end{figure*}

The same straightforward explanation can not be trivially repeated
when one considers situations conducive to the electron firehose instability
(EFHI) acting on the $L$ mode, such as those shown in Fig. \ref{fig:disp_EFHI_cutoff}\@. 
In this case, it was necessary to adopt a core + halo model for the
ions as well, where the core is described by a Maxwellian, whereas
the ion halo is given by an RKD\@. Both distributions are isotropic
and the RKD parameters for the ions are $\kappa_{i}=1$ and $\mu_{i}=0.01$.
The parameters considered here are similar to those adopted in Fig.
7 of Ref. \onlinecite{Schroeder+25/03} and the physical parameters
for all species $\left(e,i\right)$ and populations $\left(c,h\right)$
are clearly identified by the labels.

In Fig.~\ref{fig:disp_EFHI_cutoff}, the reference case is $\mu_{\perp e,h}=\mu_{\parallel e,h}=0.01$, which is again depicted using black line and dots.
The sequence of plots
black $\to$ green $\to$ magenta $\to$ purple $\to$ red is obtained
by a steadfast increase of $\mu_{\parallel e,h}$ keeping $\mu_{\perp e,h}$
fixed
(\emph{i.e.}, an increase in $r_{Te,h}$). 
The observed consequence is a rightshift in both the dispersion
relation and the growth rates. The former depicting a narrow range
of almost zero group velocity, whereas the latter shifts keeping the
maximum growth rate approximately constant, with a concomitant increase
in the unstable spectral range.

A visible increase in the growth rate is only shown by the cyan results,
where $\mu_{\perp e,h}=0.05$ was increased from the reference, keeping
$\mu_{\parallel e,h}=0.01$\@. In this case, the increase in $\gamma$
can be interpreted as an enhancement of the the effective anisotropy
that gives rise to the electron firehose instability,
which occurs when $T_{\perp e,h} < T_{\parallel e,h}$.

\begin{figure*}[ht!]
\includegraphics[width=1\textwidth]{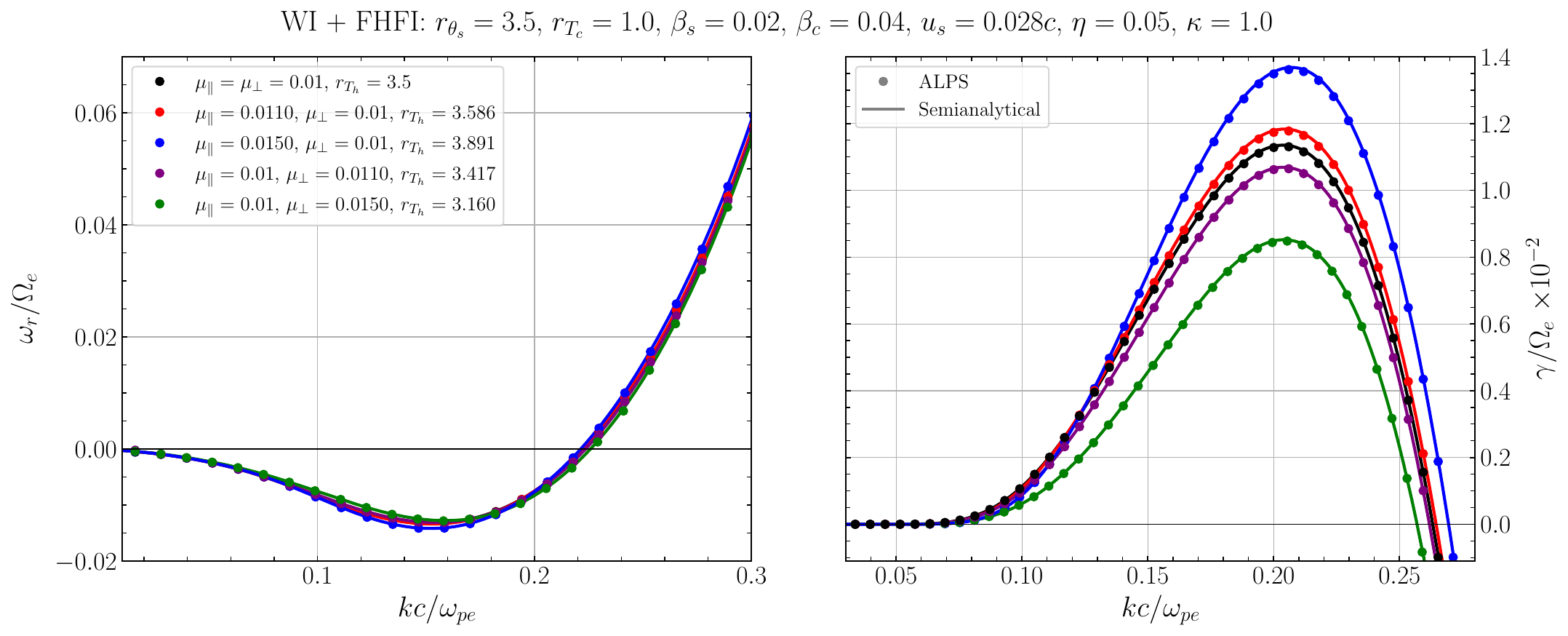}
\caption{Instability occurring in the $L$ mode due to the combined effect
of large heat flux and anisotropy $r_{\theta_{s}}>1$.
Results corresponding to the reference case $\kappa=1$ and $\mu_{\perp s}=0.01$
of Fig. \ref{fig:disp_WI_FHFI}, varying $\mu_{\perp s}\protect\neq\mu_{\parallel s}$.}
\label{fig:disp_WI_heatflux_cutoff}
\end{figure*}

The analysis is concluded by returning to the standard situation of
a single Maxwellian ion species and a core + strahl combination for
the electrons, considering now two cases of instabilities excited
by the interplay of temperature anisotropies with heat flux in the
strahl.
Figure \ref{fig:disp_WI_heatflux_cutoff} considers again the combination
of heat flux with $\theta_{\perp s} > \theta_{\parallel s}$ that generates instabilities in the $L$ mode. The results take the most extreme combination of parameters in Fig. \ref{fig:disp_WI_FHFI} ($\kappa_{s}=1$,
$\mu_{s}=0.01$) and considers this the reference case.
In Fig.~\ref{fig:disp_WI_FHFI}, the instability is increased as the plasma becomes more suprathermal, via the reduction of the $\kappa$ index.
On the other hand, Fig.~\ref{fig:disp_WI_heatflux_cutoff} shows that increasing
$\mu_{\parallel s}$ while keeping $\mu_{\perp s}$ constant (plots
red and blue) also results in an increase of the instability, which can be interpreted as an enhancement of the effective parallel temperature, compared to the perpendicular temperature. The same behavior is observed with the pure WI case shown in Fig.~\ref{fig:disp_WI_cutoff}, but with the difference that now the instability occurs in the $L$
mode.
%
%
The opposite happens when $\mu_{\perp s}$ increases keeping $\mu_{\parallel s}$
fixed. The instability is increasingly suppressed in this situation.

\begin{figure*}[ht!]
\includegraphics[width=1\textwidth]{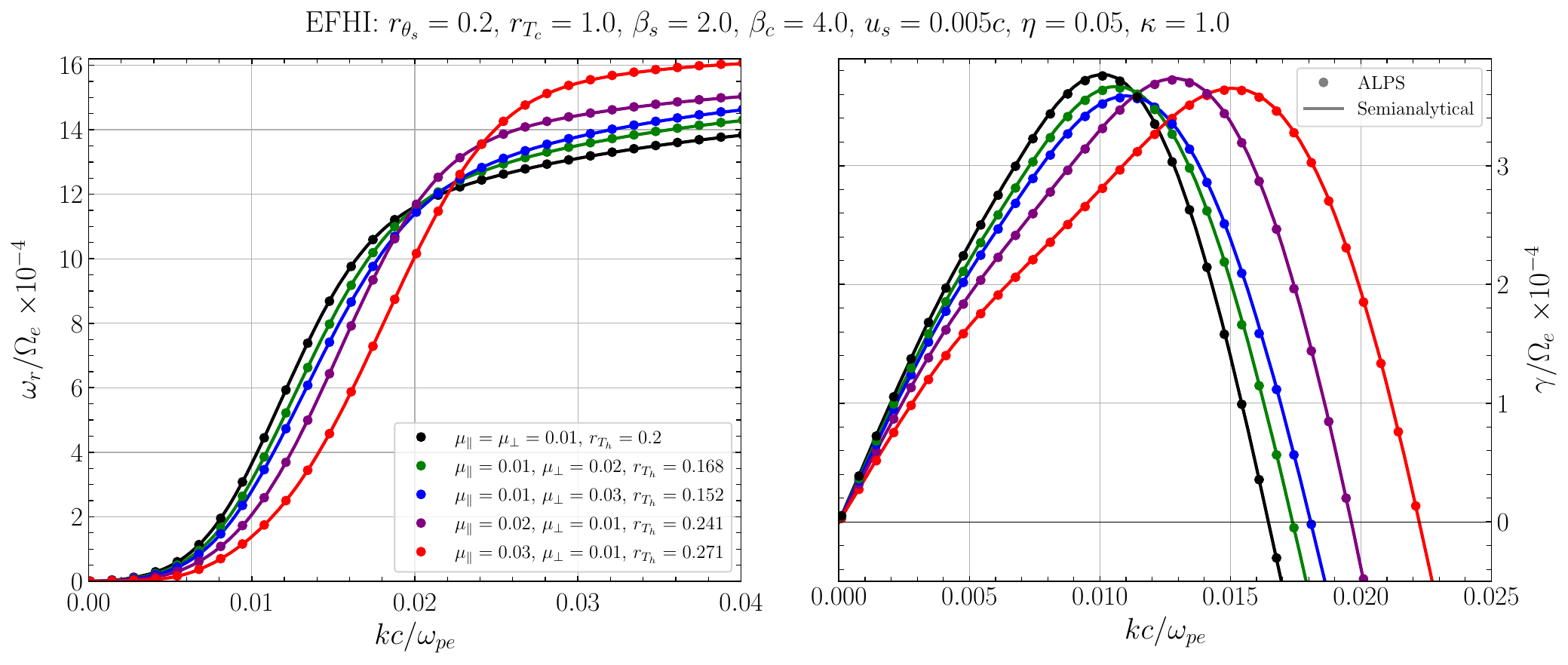}
\caption{Instability occurring in the $L$ mode due to the combined effect
of moderate heat flux and anisotropy $r_{\theta_{s}}<1$. Results corresponding to the reference case $\kappa=1$
and $\mu_{\perp s}=0.01$ of Fig. \ref{fig:disp_EFHI}, varying $\mu_{\perp s}\protect\neq\mu_{\parallel s}$.}
\label{fig:disp_EFHI_heatflux_cutoff}
\end{figure*}

Finally, Fig. \ref{fig:disp_EFHI_heatflux_cutoff} takes the parameters
$\kappa_{s}=1$ and $\mu_{s}=0.01$ of Fig. \ref{fig:disp_EFHI},
using them as the reference case for the excitation of the instability
in the $L$ mode occurring due to moderate heat flux combined with
a $\theta_{\perp s} < \theta_{\parallel s}$ anisotropy.
The results show that the instability shuffles around the reference
case for different anisotropies in the regularization parameter, with
an apparent tendency of slowly shifting the peak to higher wavenumber
whilst remaining largely at the same level, regardless of the anisotropy
in the regularization parameter.

The results presented in Secs. \ref{sec:Semi-anisotropic} and \ref{sec:Fully-anisotropic}
were basically intended to demonstrate initial applications of the
formalism developed in this work and by no means cover the full parameter
space relevant to the considered instabilities. The most important
outcome of the cases studied here is - based on the mutual validation provided
by the excellent agreement between the results obtained with fully numerical treatment (employing
the ALPS code) and those derived with the newly developed theoretical semi-analytical formalism - the new model describing the wave propagation and wave-particle interaction in fully anisotropic
regularized kappa plasmas.

\section{Summary and conclusions}\label{sec:Summary}

In the present paper we provide a theoretical framework for studying wave instabilities in plasmas described by fully anisotropic RKDs, extending the (semi-)analytical foundations available for these type of distributions. We have defined new plasma dispersion functions for plasmas modelled by the fully anisotropic RKD, where the distribution's anisotropy manifests itself not only in different thermal velocity spreads between the parallel and perpendicular directions relative to the ambient magnetic field, but also in different cut-offs in these directions.
Several important mathematical properties of these new dispersion functions have been derived, providing analytical expressions for the dielectric tensor components and the corresponding dispersion equations for both transversal and longitudinal waves propagating parallel to the magnetic field.
The dispersion equations for transverse, circularly-polarized parallel-propagating waves were numerically solved using two different approaches: semi-analytically, from the newly derived expressions in the present work and fully numerically by employing the dispersion solver ALPS\@.  The relevant equations were solved employing realistic core + strahl and core + halo electron models, with physical parameters similar to those observed \emph{in situ} in the solar wind, where both temperature anisotropy and heat-flux instabilities are observed. The dispersion relations and the growth rates obtained by either approach displayed excellent agreement in all adopted parameter sets, thus providing a mutual validation for both techniques and thereby demonstrating 
that the present formalism constitutes a reliable and versatile tool for the study of wave instabilities in plasmas described by anisotropic RKDs.

The results obtained when the distribution of suprathermal electrons featured both kinds of anisotropy, i.e.\ in the thermal velocity spread parameters and in the regularization parameters, demonstrate how such regularization can, depending on the nature of the instability and on the relative difference between the parallel and perpendicular cut-offs, act either as a stabilization that suppresses unstable modes or provide an additional source of free energy, thus enhancing the growth rates. These results indicate that the regularization parameters are not merely mathematical tools used to control the high-energy tails and the effect of superluminal particles (always present in a non-relativisitic treatment) of the distribution, but represent physically well-motivated quantities capable of influencing wave-particle interactions and energy transport.

The above findings overcome the inherent limitations of SKDs and significantly extend previous studies that were restricted to either purely numerical studies of anisotropic RKDs or semi-analytical derivations of isotropic RKDs.  
Given the widespread occurrence of suprathermal populations in various space plasma environments, such as the solar corona, solar wind, or planetary magnetospheres, our results contribute to a more realistic physical description of wave-particle interaction and wave propagation in these environments and provide valuable benchmarks for future numerical, quasi-linear as well as observational studies.

\begin{acknowledgments}
We gratefully acknowledge financial support from DFG for a Mercator Fellowship for R.G., who also acknowledges support provided by Conselho Nacional de Desenvolvimento Científico e Tecnológico (CNPq), Grant No. 313330/2021-2. The research was funded within the WEAVE scheme by the Deutsche Forschungsgesellschaft (DFG), Project No. FI 706/31-1, and the Belgian FWO-Vlaanderen, project G002523N. The authors acknowledge the use of the ALPS code. \end{acknowledgments}

\appendix

\section{Definitions and properties of integrals and special functions}\label{sec:Definitions-and-properties}

In this appendix, we discuss some representations for the frequently-appearing
integral $I_{\beta}^{\left(p_{1},p_{2}\right)}\left(\mu_{\parallel},\mu_{\perp},z\right)$,
defined by 
\begin{equation}
I_{\beta}^{\left(p_{1},p_{2}\right)}=\int_{0}^{\infty}\frac{\beta^{\zeta-1}e^{-\left(1+z^{2}/\eta\right)\beta}d\beta}{\left(\mu_{\parallel}+\beta/\eta\right)^{p_{1}}\left(\mu_{\perp}+\beta/\eta\right)^{p_{2}}}.\label{eq:RKAP3:I_beta-gen-def}
\end{equation}

Equivalent forms for \eqref{eq:RKAP3:I_beta-gen-def} can be obtained
depending on whether $\mu_{\perp}>\mu_{\parallel}$ or $\mu_{\perp}<\mu_{\parallel}$
and whether $z=0$ or $z\neq0$\@. Here, we derive generic expressions
for the case $z=0$, whereas for the case $z\neq0$ we only need the
cases $p_{1}=-1,0$.

\subsection{Case $z=0$}\label{subsec:Case-z0}

\begin{subequations}
\label{eq:RKAP2:I_beta-gen-def}

This is the simplest case, when \eqref{eq:RKAP3:I_beta-gen-def} reduces
to 
\begin{equation}
I_{\beta}^{\left(p_{1},p_{2}\right)}\left(0\right)=\int_{0}^{\infty}\frac{\beta^{\zeta-1}e^{-\beta}d\beta}{\left(\mu_{\parallel}+\beta/\eta\right)^{p_{1}}\left(\mu_{\perp}+\beta/\eta\right)^{p_{2}}}.\label{eq:RKAP2:IBGD-1}
\end{equation}

If we transform the integration variable to $\beta=\eta\mu_{\perp}t$,
then 
\begin{equation}
I_{\beta}^{\left(p_{1},p_{2}\right)}\left(0\right)=\frac{\left(\eta\mu_{\perp}\right)^{\zeta}}{\mu_{\perp}^{p_{1}+p_{2}}}\int_{0}^{\infty}dt\,\frac{t^{\zeta-1}e^{-\eta\mu_{\perp}t}}{\left(1+t-\mathcal{A}_{\boldsymbol{\mu}}\right)^{p_{1}}\left(1+t\right)^{p_{2}}}.\label{eq:RKAP2:IBGD-2}
\end{equation}

On the other hand, if we define $\beta=\eta\mu_{\parallel}t$, then
\begin{equation}
I_{\beta}^{\left(p_{1},p_{2}\right)}\left(0\right)=\frac{\left(\eta\mu_{\parallel}\right)^{\zeta}}{\mu_{\parallel}^{p_{1}+p_{2}}}\int_{0}^{\infty}dt\,\frac{t^{\zeta-1}e^{-\eta\mu_{\parallel}t}}{\left(1+t\right)^{p_{1}}\left(1+t-A_{\boldsymbol{\mu}}\right)^{p_{2}}}.\label{eq:RKAP2:IBGD-3}
\end{equation}
\end{subequations}

In (\ref{eq:RKAP2:I_beta-gen-def}a, b) we have defined the anisotropy
parameters 
\begin{equation}
\begin{aligned}A_{\boldsymbol{\mu}} & =1-\frac{\mu_{\perp}}{\mu_{\parallel}}, & \mathcal{A}_{\boldsymbol{\mu}} & =1-\frac{\mu_{\parallel}}{\mu_{\perp}},\end{aligned}
\label{eq:Anisotropy-pars}
\end{equation}
which are related by 
\begin{align*}
\mathcal{A}_{\boldsymbol{\mu}} & =\frac{A_{\boldsymbol{\mu}}}{A_{\boldsymbol{\mu}}-1}, & A_{\boldsymbol{\mu}} & =\frac{\mathcal{A}_{\boldsymbol{\mu}}}{\mathcal{A}_{\boldsymbol{\mu}}-1}.
\end{align*}

The integral $I_{\beta}^{\left(p_{1},p_{2}\right)}\left(\mu_{\parallel},\mu_{\perp},0\right)$
can be evaluated via numerical quadrature routines, but additional
representations are possible. We start by noticing that when the regularization
parameters are isotropic $\left(\mu_{\parallel}=\mu_{\perp}=\mu\right)$,
\eqref{eq:RKAP2:IBGD-2} reduces to 
\[
I_{\beta}^{\left(p_{1},p_{2}\right)}\left(\mu\right)=\frac{\left(\eta\mu\right)^{\zeta}}{\mu^{p_{1}+p_{2}}}\int_{0}^{\infty}dt\,\frac{t^{\zeta-1}e^{-\eta\mu t}}{\left(1+t\right)^{p_{1}+p_{2}}}.
\]
Given the definition of the Tricomi function $U\left(a,b,z\right)$,\cite{Daalhuis-NIST10a}
\begin{equation}
U\left(a,b,z\right)=\frac{1}{\Gamma\left(a\right)}\int_{0}^{\infty}dt\,t^{a-1}\left(1+t\right)^{b-a-1}e^{-zt},\label{eq:RKAP2:Tricomi-Integ_rep-1}
\end{equation}
valid when $\Re a>0$ and $\left|\arg\left(z\right)\right|<\nicefrac{\pi}{2}$,
we obtain 
\begin{equation}
I_{\beta}^{\left(p_{1},p_{2}\right)}\left(\mu\right)=\frac{\left(\eta\mu\right)^{\zeta}}{\mu^{p_{1}+p_{2}}}\Gamma\left(\zeta\right)U\left(\zeta,\zeta+1-p_{1}-p_{2},\eta\mu\right).\label{eq:RKAP2:I_beta-gen-isotropic}
\end{equation}

\begin{widetext}
\begin{subequations}
\label{eq:RKAP2:I_beta-gen-series}

Now, in the anisotropic case, we can derive power series expansions,
valid when the anisotropy parameters are small. When $\left|\mathcal{A}_{\boldsymbol{\mu}}\right|<1$,
we can employ the binomial theorem and expand in \eqref{eq:RKAP2:IBGD-2},
\[
\frac{1}{\left(1+t-\mathcal{A}_{\boldsymbol{\mu}}\right)^{p_{1}}}=\frac{1}{\left(1+t\right)^{p_{1}}}\sum_{\ell=0}^{\infty}\frac{\left(p_{1}\right)_{\ell}}{\ell!}\left(\frac{\mathcal{A}_{\boldsymbol{\mu}}}{1+t}\right)^{\ell},
\]
In this case, we obtain 
\begin{equation}
I_{\beta}^{\left(p_{1},p_{2}\right)}\left(\mu_{\parallel},\mu_{\perp},0\right)=\frac{\left(\eta\mu_{\perp}\right)^{\zeta}}{\mu_{\perp}^{p_{1}+p_{2}}}\Gamma\left(\zeta\right)\sum_{\ell=0}^{\infty}\frac{\left(p_{1}\right)_{\ell}}{\ell!}U\left(\zeta,\zeta+1-p_{1}-p_{2}-\ell,\eta\mu_{\perp}\right)\mathcal{A}_{\boldsymbol{\mu}}^{\ell},
\end{equation}
where we have identified the Tricomi function in \eqref{eq:RKAP2:Tricomi-Integ_rep-1}
again.

On the other hand, when $\left|A_{\boldsymbol{\mu}}\right|<1$, we
can proceed in the same fashion with \eqref{eq:RKAP2:IBGD-3} and
obtain the alternative expansion 
\begin{equation}
I_{\beta}^{\left(p_{1},p_{2}\right)}\left(\mu_{\parallel},\mu_{\perp},0\right)=\frac{\left(\eta\mu_{\parallel}\right)^{\zeta}}{\mu_{\parallel}^{p_{1}+p_{2}}}\Gamma\left(\zeta\right)\sum_{\ell=0}^{\infty}\frac{\left(p_{2}\right)_{\ell}}{\ell!}U\left(\zeta,\zeta+1-p_{1}-p_{2}-\ell,\eta\mu_{\parallel}\right)A_{\boldsymbol{\mu}}^{\ell}.
\end{equation}
Notice that both series reduce to \eqref{eq:RKAP2:I_beta-gen-isotropic}
in the $\mu$-isotropic limit.
\end{subequations}

The direct evaluation of $I_{\beta}^{\left(p_{1},p_{2}\right)}\left(\mu_{\parallel},\mu_{\perp}\right)$
from the definition (\ref{eq:RKAP2:I_beta-gen-def}) requires the
numerical computation of improper integrals. We can, alternatively,
obtain representations involving definite integrals over a finite
region. We start by defining the new parameter $\mathcal{U}_{\alpha,\beta}^{\left(\eta,\zeta,\boldsymbol{\mu}\right)}$
as 
\begin{equation}
\mathcal{U}_{\alpha,\beta}^{\left(\eta,\zeta,\boldsymbol{\mu}\right)}=\frac{\Gamma\left(\alpha+\beta+\nicefrac{3}{2}\right)}{\Gamma\left(\alpha+\nicefrac{1}{2}\right)\Gamma\left(\beta+1\right)}\int_{0}^{1}dx\begin{cases}
{\displaystyle x^{\alpha-1/2}\left(1-x\right)^{\beta}U\left(\alpha+\beta+\frac{3}{2},\alpha+\beta+\frac{5}{2}-\zeta,\eta\mu_{\perp}\left(1-\mathcal{A}_{\boldsymbol{\mu}}x\right)\right),} & \mu_{\parallel}<\mu_{\perp}\\
{\displaystyle \left(1-x\right)^{\alpha-1/2}x^{\beta}U\left(\alpha+\beta+\frac{3}{2},\alpha+\beta+\frac{5}{2}-\zeta,\eta\mu_{\parallel}\left(1-A_{\boldsymbol{\mu}}x\right)\right),} & \mu_{\parallel}>\mu_{\perp},
\end{cases}\label{eq:RKAP3:CalU-general}
\end{equation}
which is valid for $\alpha > -\nicefrac{1}{2}$ and $\beta > -1$, and where $U\left(a,b,z\right)$ is again the Tricomi function \eqref{eq:RKAP2:Tricomi-Integ_rep-1}\@.
Definition \eqref{eq:RKAP3:CalU-general} requires now the evaluation
of Euler-type integrals.

It can be shown that the integral $I_{\beta}^{\left(p_{1},p_{2}\right)}\left(\mu_{\parallel},\mu_{\perp}\right)$
and the parameter $\mathcal{U}_{\alpha,\beta}^{\left(\eta,\zeta,\boldsymbol{\mu}\right)}$
are related by the identity 
\begin{equation}
I_{\beta}^{\left(p_{1},p_{2}\right)}\left(\mu_{\parallel},\mu_{\perp},0\right)=\Gamma\left(\zeta\right)\eta^{p_{1}+p_{2}}\mathcal{U}_{p_{1}-\nicefrac{1}{2},p_{2}-1}^{\left(\eta,\zeta,\boldsymbol{\mu}\right)},\label{eq:RKAP3:I_beta-gen-CalU-gen}
\end{equation}
which is valid in both cases $\mu_{\parallel}<\mu_{\perp}$ and $\mu_{\parallel}>\mu_{\perp}$\@.

In order to derive \eqref{eq:RKAP3:I_beta-gen-CalU-gen}, we introduce
into \eqref{eq:RKAP2:IBGD-1} the identity 
\[
\frac{1}{A_{1}^{\alpha_{1}}A_{2}^{\alpha_{2}}}=\frac{\Gamma\left(\alpha_{1}+\alpha_{2}\right)}{\Gamma\left(\alpha_{1}\right)\Gamma\left(\alpha_{2}\right)}\int_{0}^{1}dx_{1}\int_{0}^{1}dx_{2}\frac{x_{1}^{\alpha_{1}-1}x_{2}^{\alpha_{2}-1}\delta\left(1-x_{1}-x_{2}\right)}{\left(x_{1}A_{1}+x_{2}A_{2}\right)^{\alpha_{1}+\alpha_{2}}},
\]
known as Feynman parametrization, and which is valid for $\alpha_{1},\alpha_{2}>0$\@.
Restricting ourselves first to the particular case $0<\mathcal{A}_{\boldsymbol{\mu}}<1$
$\left(\mu_{\parallel}<\mu_{\perp}\right)$, we can easily obtain
\begin{align*}
I_{\beta}^{\left(p_{1},p_{2}\right)}\left(\mu_{\parallel},\mu_{\perp},0\right) & =\frac{\Gamma\left(p_{1}+p_{2}\right)\left(\eta\mu_{\perp}\right)^{\zeta}}{\Gamma\left(p_{1}\right)\Gamma\left(p_{2}\right)\mu_{\perp}^{p_{1}+p_{2}}}\int_{0}^{1}dx\,x^{p_{1}-1}\left(1-x\right)^{p_{2}-1}\left(1-\mathcal{A}_{\boldsymbol{\mu}}x\right)^{\zeta-p_{1}-p_{2}}\int_{0}^{\infty}dy\,\frac{y^{\zeta-1}e^{-\eta\mu_{\perp}\left(1-\mathcal{A}_{\boldsymbol{\mu}}x\right)y}}{\left(1+y\right)^{p_{1}+p_{2}}},
\end{align*}
after changing the integration variable $\beta\to y$ through $\beta=\eta\mu_{\perp}\left(1-\mathcal{A}_{\boldsymbol{\mu}}x\right)y$\@.
Comparing the last integral with \eqref{eq:RKAP2:Tricomi-Integ_rep-1}
and then comparing the result with \eqref{eq:RKAP3:CalU-general},
we obtain identity \eqref{eq:RKAP3:I_beta-gen-CalU-gen}\@. During
the demonstration, we also employed the Kummer transformation\cite{Daalhuis-NIST10a}
$U\left(a,b,z\right)=z^{1-b}U\left(a-b+1,2-b,z\right)$\@. The demonstration
for the case $\mu_{\parallel}>\mu_{\perp}$ $\left(0<A_{\boldsymbol{\mu}}<1\right)$
proceeds in a similar fashion.

Finally, identity \eqref{eq:RKAP3:I_beta-gen-CalU-gen}
provides two additional expressions valid for any values of the parameters $\alpha$ and $\beta$\@.  
Firstly, from (\ref{eq:RKAP2:I_beta-gen-series}a, b) one obtains the expansions
\begin{equation}
\mathcal{U}_{\alpha,\beta}^{\left(\eta,\zeta,\boldsymbol{\mu}\right)}=\begin{cases}
{\displaystyle \left(\eta\mu_{\perp}\right)^{\zeta-\alpha-\beta-3/2}\sum_{\ell=0}^{\infty}\frac{\left(\alpha+\nicefrac{1}{2}\right)_{\ell}}{\ell!}U\left(\zeta,\zeta-\alpha-\beta-\frac{1}{2}-\ell,\eta\mu_{\perp}\right)\mathcal{A}_{\boldsymbol{\mu}}^{\ell},} & \left|\mathcal{A}_{\boldsymbol{\mu}}\right|<1\\
{\displaystyle \left(\eta\mu_{\parallel}\right)^{\zeta-\alpha-\beta-3/2}\sum_{\ell=0}^{\infty}\frac{\left(\beta+1\right)_{\ell}}{\ell!}U\left(\zeta,\zeta-\alpha-\beta-\frac{1}{2}-\ell,\eta\mu_{\parallel}\right)A_{\boldsymbol{\mu}}^{\ell},} & \left|A_{\boldsymbol{\mu}}\right|<1.
\end{cases}\label{eq:RKAP3:CalU-gen-series}
\end{equation}

Secondly, from (\ref{eq:RKAP2:IBGD-1}) we have the direct evaluation from the improper integral
\begin{displaymath}
\mathcal{U}^{\left(\eta,\zeta,\boldsymbol{\mu}\right)}_{\alpha_{1},\alpha_{2}}=\frac{1}{\Gamma\left(\zeta\right)}\eta^{-\left(\alpha_{1}+\alpha_{2}+3/2\right)}\int^{\infty}_{0}\frac{d\beta\,\beta^{\zeta-1}e^{-\beta}}{\left(\mu_{\parallel}+\beta/\eta\right)^{\alpha_{1}+\nicefrac{1}{2}}\left(\mu_{\perp}+\beta/\eta\right)^{\alpha_{2}+1}}.
\end{displaymath}

Important particular cases of $\mathcal{U}_{\alpha,\beta}^{\left(\eta,\zeta,\boldsymbol{\mu}\right)}$ can be also derived.  The first one is the $\mu$-isotropic limit $(\mu_\perp = \mu_\parallel = \mu)$, which can be readily obtained from (\ref{eq:RKAP3:CalU-general}) or (\ref{eq:RKAP3:CalU-gen-series}), resulting 
\begin{equation}
\mathcal{U}^{\left(\eta,\zeta,\mu\right)}_{\alpha,\beta}=U\left(\alpha+\beta+\frac{3}{2},\alpha+\beta+\frac{5}{2}-\zeta,\eta\mu\right) = \left(\eta\mu\right)^{\zeta-\alpha-\beta-3/2}U\left(\zeta,\zeta-\alpha-\beta-\frac{1}{2},\eta\mu\right).
\label{eq:CalU-Isotropic_mu}
\end{equation}

Another important case is the limit $\mu_{\perp,\parallel} \to 0$, in which case $\mathcal{U}_{\alpha,\beta}^{\left(\eta,\zeta,\boldsymbol{\mu}\right)}$ reduces to the corresponding expression for the standard bi-Kappa distribution.  This is accomplished by taking the adequate limit of $U(a,b,z)$ when $z\to 0$ that will introduce the constraint on the lower value of $\kappa$ imposed by the evaluation of a given moment of the VDF\@.  Following the procedure outlined in Refs. \onlinecite{Gaelzer+24/07} or \onlinecite{Gaelzer-Ziebell-2025}, expression (\ref{eq:RKAP3:CalU-general}) provides the limiting form
\begin{equation}
\mathcal{U}^{\left(\eta,\zeta,0\right)}_{\alpha,\beta} = \frac{\Gamma\left(\zeta-\alpha-\beta-\nicefrac{3}{2}\right)}{\Gamma\left(\zeta\right)}\quad\left(\zeta > \frac{3}{2}+\alpha+\beta\right).
\label{eq:RKAP3:CalU-SKD_limit}
\end{equation}

\subsection{Case $z\protect\neq0$}

\begin{subequations}
\label{eq:RKAP3:I_beta-gen-z}

Returning to \eqref{eq:RKAP3:I_beta-gen-def}, we will only need the
particular cases $p_{1}=-1,0$, for which we will obtain closed-form
expression by first changing the integration variable with $\beta=\eta\mu_{\perp}t$
and then comparing with \eqref{eq:RKAP2:Tricomi-Integ_rep-1}\@.
Accordingly, we obtain 
\begin{align}
I_{\beta}^{\left(0,p_{2}\right)}\left(\mu_{\parallel},\mu_{\perp},z\right) & =\Gamma\left(\zeta\right)\frac{\left(\eta\mu_{\perp}\right)^{\zeta}}{\mu_{\perp}^{p_{2}}}U\left(\zeta,\zeta+1-p_{2},\eta\mu_{\perp}\left(1+\frac{z^{2}}{\eta}\right)\right)\label{eq:RKAP3:I_beta-gen-p10-1}\\
I_{\beta}^{\left(-1,p_{2}\right)}\left(\mu_{\parallel},\mu_{\perp},z\right) & =\Gamma\left(\zeta\right)\frac{\left(\eta\mu_{\perp}\right)^{\zeta}}{\mu_{\perp}^{p_{2}-1}}\left[U\left(\zeta,\zeta+2-p_{2},\eta\mu_{\perp}\left(1+\frac{z^{2}}{\eta}\right)\right)-\mathcal{A}_{\boldsymbol{\mu}}U\left(\zeta,\zeta+1-p_{2},\eta\mu_{\perp}\left(1+\frac{z^{2}}{\eta}\right)\right)\right].\label{eq:RKAP3:I_beta-gen-p1m1-1}
\end{align}
\end{subequations}
\end{widetext}

\section{Susceptibility tensors for drifting bi-Maxwellian and bi-kappa populations}\label{sec:bMd-bSKd}

\begin{subequations}
\label{eq:ST-bMd}

The specific analytical expressions of the susceptibility tensor components
(\ref{eq:RKAP2:DT-magnetized-parallel}a-c) for the 
dbM VDF given by (\ref{eq:Maxwellian-VDF})
can be found in several textbooks of kinetic theory of plasmas (see,
\emph{e.g.}, Ref. \onlinecite{Brambilla98}) and is given by
\begin{align}
\chi_{1}^{\text{\rm dbM}} & =\frac{1}{2}\frac{\omega_{pa}^{2}}{\omega^{2}}\sum_{s=\pm1}\left[\varsigma_{0}Z\left(\varsigma_{s}\right)+\frac{1}{2}A_{\boldsymbol{\theta}}Z^{\prime}\left(\varsigma_{s}\right)\right]\\
\chi_{2}^{\text{\rm dbM}} & =\frac{i}{2}\frac{\omega_{pa}^{2}}{\omega^{2}}\sum_{s=\pm1}s\left[\varsigma_{0}Z\left(\varsigma_{s}\right)+\frac{1}{2}A_{\boldsymbol{\theta}}Z^{\prime}\left(\varsigma_{s}\right)\right]\\
\chi_{3}^{\text{\rm dbM}} & =-\frac{\omega_{pa}^{2}}{\omega^{2}}\xi_{0}^{2}Z^{\prime}\left(\varsigma_{0}\right),
\end{align}
where 
\[
Z\left(\psi\right)=\frac{1}{\sqrt{\pi}}\int_{-\infty}^{\infty}\frac{e^{-y^{2}}dy}{y-\psi}\qquad\left(\Im\psi>0\right)
\]
is the Fried \& Conte function\cite{FriedConte61} and $Z^{\prime}\left(\psi\right)$
its derivative.

Here, the parameters are 
\begin{align*}
\varsigma_{n} & =\xi_{n}-\frac{U_{\parallel}}{\theta_{\parallel}}, & \xi_{n} & =\frac{\omega-n\Omega}{k_{\parallel}\theta_{\parallel}}, & A_{\boldsymbol{\theta}} & =1-\frac{T_{M,\perp}}{T_{M,\parallel}} = 1- {\theta^2_\perp \over \theta^2_\parallel},
\end{align*}
where $T_{M,\perp(\parallel)}$ is the perpendicular (parallel) temperature
and $\theta_{\perp(\parallel)}$ are the corresponding components of the
thermal velocity.
\end{subequations}

The susceptibility tensor for a drifting standard Kappa distribution
can be obtained as a special case of the general expressions derived
in Refs. \onlinecite{Gaelzer+16/06} and \onlinecite{Meneses+18/11}\@.
The drifting standard bi-Kappa distribution function (dbSKD) adopted here
is
\begin{equation}
f_{\mathrm{dbSKD}}\left(v_{\parallel},v_{\perp}\right)=\frac{g\left(\kappa\right)}{\pi^{3/2}\theta_{\parallel}\theta_{\perp}^{2}}\left[1+\frac{\left(v_{\parallel}-U_{\parallel}\right)^{2}}{\kappa \theta_{\parallel}^{2}}+\frac{v_{\perp}^{2}}{\kappa \theta_{\perp}^{2}}\right]^{-\left(\kappa+1\right)},\label{eq:KAP6:KappaDistFunction}
\end{equation}
 which is valid for $\kappa>\nicefrac{1}{2}$ and where $g\left(\kappa\right)=\Gamma\left(\kappa\right)/\kappa^{1/2}\Gamma\left(\kappa-\nicefrac{1}{2}\right)$\@.

\begin{subequations}
\label{eq:ST-bSKd}

For the distribution (\ref{eq:KAP6:KappaDistFunction}), the susceptibility
tensor is 
\begin{align}
\chi_{1}^{\mathrm{(dbSKD)}} & =\frac{1}{2}\frac{\omega_{pa}^{2}}{\omega^{2}}\sum_{s=\pm1}\left[\varsigma_{0}Z_{\kappa}^{\left(0\right)}\left(\varsigma_{s}\right)+\frac{1}{2}A_{\boldsymbol{\theta}}Z_{\kappa}^{\left(-1\right)\prime}\left(\varsigma_{s}\right)\right]\\
\chi_{2}^{\mathrm{(dbSKD)}} & =\frac{i}{2}\frac{\omega_{pa}^{2}}{\omega^{2}}\sum_{s=\pm1}s\left[\varsigma_{0}Z_{\kappa}^{\left(0\right)}\left(\varsigma_{s}\right)+\frac{1}{2}A_{\boldsymbol{\theta}}Z_{\kappa}^{\left(-1\right)\prime}\left(\varsigma_{s}\right)\right]\\
\chi_{3}^{\mathrm{(dbSKD)}} & =-\frac{\omega_{pa}^{2}}{k_{\parallel}^{2}\theta_{\parallel}^{2}}Z_{\kappa}^{\left(0\right)\prime}\left(\varsigma_{0}\right).
\end{align}
%
%
\end{subequations}

In (\ref{eq:ST-bSKd}a-c), the tensor components are written in terms
of the (standard) kappa dispersion function 
\[
Z_{\kappa}^{\left(\beta\right)}\left(\xi\right)=\frac{\kappa^{-\beta-1/2}\Gamma\left(\kappa+\beta\right)}{\pi^{1/2}\Gamma\left(\kappa-\nicefrac{1}{2}\right)}\int_{-\infty}^{\infty}ds\frac{\left(1+s^{2}/\kappa\right)^{-\left(\kappa+\beta\right)}}{s-\xi},
\]
which is valid for $\kappa+\beta>0$\@. This function and its derivative
can be evaluated from the representation 
\begin{multline*}
Z_{\kappa}^{\left(\beta\right)}\left(\xi\right)=i\frac{\kappa^{-\beta-1/2}\Gamma\left(\kappa+\beta+\nicefrac{1}{2}\right)}{\left(\kappa+\beta\right)\Gamma\left(\kappa-\nicefrac{1}{2}\right)}\\
\times F\left[{1,2\left(\kappa+\beta\right)\atop \kappa+\beta+1};\frac{1}{2}\left(1+\frac{i\xi}{\kappa^{1/2}}\right)\right],
\end{multline*}
where $F\left(\cdots\right)$ is the Gauss hypergeometric function.\cite{Daalhuis-NIST10b}

The function $Z_{\kappa}^{\left(\beta\right)}\left(\xi\right)$ and
its derivative are also related by 
\[
Z_{\kappa}^{\left(\beta\right)\prime}\left(\xi\right)=-2\left[\frac{\Gamma\left(\kappa+\beta+\nicefrac{1}{2}\right)}{\kappa^{\beta+1}\Gamma\left(\kappa-\nicefrac{1}{2}\right)}+\xi Z_{\kappa}^{\left(\beta+1\right)}\left(\xi\right)\right].
\]

\bibliographystyle{aipnum4-1}
\bibliography{RKAP_Fully_Anisotropic-1}

\end{document}